\documentclass[preprint,5p,times, authoryear, sort, comma, singleblind, nopreprintline]{elsarticle}
\usepackage[utf8]{inputenc}
\usepackage{amsmath,amsfonts,amssymb, mathtools, blindtext}
\usepackage{natbib}
\usepackage{multirow}
\usepackage{subfig}
\usepackage{graphicx}
\usepackage{tabularx}
\usepackage{amssymb}
\usepackage{float}
 \usepackage{csvsimple} 
\usepackage{xcolor}
\usepackage{appendix}
\usepackage{array, makecell}
\usepackage{adjustbox}
\usepackage[figuresright]{rotating}

\graphicspath{ {./experiment-figs/} }

\begin{document}

\begin{frontmatter}
\title{Enhancing Supply Chain Visibility with Generative AI: An Exploratory Case Study on Relationship Prediction in Knowledge Graphs}

\author[inst1]{Ge Zheng}
\author[inst1,inst2]{Alexandra Brintrup\corref{cor1}}
\affiliation[inst1]{organization={Institute for Manufacturing, University of Cambridge},country={United Kingdom}}
\affiliation[inst2]{organization={Alan Turing Institute, London},country={United Kingdom}}
\cortext[cor1]{Corresponding author: ab702@cam.ac.uk}

\begin{abstract}

A key stumbling block in effective supply chain risk management for companies and policymakers is a lack of visibility on interdependent supply network relationships. 
Relationship prediction, also called link prediction is an emergent area of supply chain surveillance research that aims to increase the visibility of supply chains using data-driven techniques. 
Existing methods have been successful for predicting relationships but struggle to extract the context in which these relationships are embedded - such as the products being supplied or locations they are supplied from. 
Lack of context prevents practitioners from distinguishing transactional relations from established supply chain relations, hindering accurate estimations of risk. 
In this work, we develop a new Generative Artificial Intelligence (Gen AI) enhanced machine learning framework that leverages pre-trained language models as embedding models combined with machine learning models to predict supply chain relationships within knowledge graphs. 
By integrating Generative AI techniques, our approach captures the nuanced semantic relationships between entities, thereby improving supply chain visibility and facilitating more precise risk management. 
Using data from a real case study, we show that GenAI-enhanced link prediction surpasses all benchmarks, and demonstrate how GenAI models can be explored and effectively used in supply chain risk management.   \\
\newline SDG 9: Industry, innovation and infrastructure

\end{abstract}


\end{frontmatter}


\section{Introduction} \label{sec: introduction}

Global supply chains emerge as companies buy products from one another to produce and deliver their own \citep{bellamy2013network}. 
They play a critical role in almost every aspect of our daily lives. 
80\% of global trade flows through multinational corporations, and one in five jobs worldwide is tied to global supply chains. 
Increased volatility and geopolitical tension in recent years have shown how vulnerable we are to supply chain disruptions, with major shortages impacting our food, medicines and supply of electric batteries. 
In tandem there is rising awareness on the exposure of global supply chains to human rights violations and unsustainable environmental practices, with US and European policy makers proposing legislative measures that demand comprehensive supply chain traceability \citep{kublbock2013eu}.

One of the key stumbling blocks to begin addressing these concerns is a lack of knowledge on interdependent supply chain connections. 
Most companies have limited visibility beyond their direct connections. 
Increasing visibility in supply chains have been a rich area of research in the past decade, with multiple technical innovations having been proposed, such as electronic product codes, radio frequency identification, and blockchain technologies. 
While these have been successful to some extent, their reach is typically limited to one or two tiers at most. 
That is because in order for tracking technology to be adopted, companies need to be willing to share data. 
There is little incentive for companies to share data on whom they buy from, for various reasons. 
Companies typically view their own supply chains as a competitive advantage, and fear that disclosing information could result in their buyers working directly with their suppliers, reveal their pricing structure, or they may simply be not wish their manufacturing and purchasing practices to be known to the buyer.

More recently, researchers have proposed a new solution to this problem, which is to use data driven methods to ``estimate'' who supplies whom, rather than rely on the willingness of companies to share data. 
Termed as ``Digital Supply Chain Surveillance'' \citep{brintrup2023digital}, these methods include network reconstruction \citep{mungo2023reconstructing}, web scraping to recognise supply relationships in text obtained from news articles and company annual reports \citep{wichmann2018towards}, and machine learning methods for predicting relationships (formally, link prediction).

Most current methods focus on a single type of relationship, such as firm-level networks that map supply or buy relationships between firms \citep{brintrup2018predicting, mungo2023reconstructing}. 
While these approaches provide valuable insights, they offer a limited understanding of supply chains due to not considering the multifaceted interactions and dependencies that exist between different entities, thereby restricting a comprehensive understanding of the entire network. 

Considering the supply chain as an interconnected network of entities and relationships, we can construct supply chain relevant data into a supply chain knowledge graph that can capture complex relationships and attributes associated with supply chain entities. 
The multiple types of relationships in the supply chain knowledge graph such as manufacturing processes required to produce a product, product flows, and types of partnerships can contribute to supply chain visibility for a comprehensive understanding of supply chain dynamics. 
It also can reveal hidden patterns, identify potential bottlenecks, and support the development of strategies to enhance network resilience.


Generative Artificial Intelligence (GenAI), a branch of machine learning, is designed to create new content, ideas, or data (known as synthetic data), by learning patterns from existing data. 
Unlike traditional Artificial Intelligence (AI) where the output depends on the given inputs, GenAIs can generate novel outputs. 
Examples could be generative machine learning models used for synthetic data generation \citep{zhang2018generative} and large language models like ChatGPT \citep{openai2024release} and Copilot \citep{coploi2023genai}, Gemini \citep{gemini2023genai} and LLaMA \citep{touvron2023llama} used for the generations of human-like text, images, audio and even videos.

GenAI as a powerful tool has gained tremendous attention in recent years and been used in various fields. 
For instance, \citep{zholus2024bindgpt} explored how a language model can be used to accurately create molecular structures, facilitating the drug discovery process. 
\citep{zhao2024advancing} introduced self-attention Generative Adversarial Networks (SAGANs) model to generate the synthetic data for solving data imbalanced issue in financial transaction data and then used it for credit card fraud detection. 
\citep{gayam2023enhancing} investigated how GenAIs contribute to the creation of music and visual art.

In the context of supply chain operation management, GenAIs have been hypothesised to empower the human workforce, improve project management processes, and help optimize manufacturing and supply chain procedure \citep{mohammed3role}.Early adopters Walmart and Maersk integrated GenAIs into their operations to optimize pricing negotiations \citep{jackson2024generative}. 
A logistics company, C.H. Robinson is exploring GenAIs for automating the freight shipment \citep{business2024robinson} while \citep{dhl2024genai} leveraged GenAIs to power chatbots to handle customer inquiries.

These early reports on GenAIs inspire us to explore its potential in enhancing supply chain visibility, particularly by predicting relationships within supply chain knowledge graphs. 
GenAIs, including pre-trained language models but not limited to, are trained on extensive datasets containing diverse tex-based information which enables them to find relevant patterns in unstructured data such as emails, reports, contracts and social media posts. 
This ability might allow it to elicit the complex structure and patterns of relationships within networks. 
GenAIs employ sophisticated neural network architectures, such as transformers, which allows the models to handle complex, non-linear relationships. 
These architectures also use mechanisms like self-attention to capture dependencies and interactions between different parts of the data. 
In supply chain networks, where relationships between entities (e.g., suppliers, manufacturers, distributors) are intricate and multifaceted, GenAIs' ability to model these complexities has great potential to enhance supply chain visibility.

However, one area of concern of applying GenAIs into industrial applications has been so called ``hallucinations'', where the model might perceive patterns that are non-existent, creating outputs that are not factual or inaccurate \citep{huang2023survey}. We cannot simply ask  GenAI such as an LLM whether a relationship exists between Jaguar Land Rover and Aisin as we cannot guarantee that the answer to this question is not hallucinated. 
This issue can undermine the reliability of the insights gained, posing risks in critical applications like relationship modelling in supply chain networks. 
To address this challenge, a hybrid approach can be adopted: using a GenAI model as an embedding model to encode data into vectors, followed by applying machine learning models for relationship prediction. 
This strategy leverages the strengths of the GenAI model in capturing intricate patterns and semantic contexts while relying on machine learning algorithms to avoid the risks of hallucinations.

In this paper, we explore the potential of GenAIs in enhancing supply chain visibility by integrating pre-trained language models with machine learning models to predict relationships within supply chain knowledge graphs.  
We also introduce a new term, ``quintuplet'', to represent more intricate relationships within the knowledge graph. 
Unlike traditional triplets that capture a single relationship between two entities, quintuplets condense multiple triplets to provide a deeper understanding of the supply chain network. 
After transferring regular triplets based knowledge graph into quintuplets based one, we are able to generate textual descriptions passed onto a pre-trained language model to retrieve vectorised relational knowledge learned by the pre-trained language model from large amount of website information, which are then further learned using a machine learning model to predict quintuplets. 
The process thus allows us to combine structured knowledge representation by a knowledge graph, with the general knowledge base that can be used to augment the graph to make additional inference.

Combining GenAIs and Knowledge Graphs is powerful and goes beyond the state of the art in the supply chain domain, because we mitigate hallucination effects of GenAIs by restricting its use to augment structured prior data, and also allow additional contextual knowledge to arise from it, which would not have arisen by merely using knowledge graph completion methods. 
Thus our contribution extends the current state of the art in supply chain relationship prediction for visibility and also allows us to provide a use case in the application of GenAIs to supply chain management research.
We compare our method to existing benchmarks with a use case in electronic vehicle battery supply chains, and present experimental results validated with a range of pre-trained language models and machine learning methods. 
Our method surpasses all existing benchmarks in accuracy, and also yields better information.

The rest of this paper is organised as follows. 
Section~\ref{sec: literature} reviews relevant existing works in the literature, including GenAIs, and existing link prediction methods in supply chain networks. 
Section~\ref{sec: methodology} presents our proposed approach for multiple connected relationship prediction in supply chain networks, including problem definition, preliminaries, and framework explanation of the proposed approach. 
Section~\ref{sec: case study} uses a case study to evaluate the proposed approach while Section~\ref{sec: conclusion} concludes this work and explains its managerial implications and potential future work directions.

\section{Related Works} \label{sec: literature}

Generative artificial intelligence and link prediction in supply chain networks are two main topics relevant to the proposed approach in this work, reviewed below.

\subsection{Generative Artificial Intelligence}

Generative Artificial Intelligence (GenAI) refers to a class of machine learning designed to generate new content including text, images, music and even videos, by learning patterns from existing data \citep{ooi2023potential}. 
GenAI models include Generative Adversarial Networks (GANs) \citep{goodfellow2014generative}, Variational Autoencoders (VAEs) \citep{kingma2019introduction}, Transformer-based ChatGPTs \citep{openai2024release}, LLaMA \citep{touvron2023llama}, DALL-E \citep{openai2022dall} and so on. 
These models have been applied into various fields, leading to significant achievements and efficiencies. 
One of the most significant capabilities of GenAI is natural language generation. 
For example, Transformer-based models like ChatGPTs \citep{openai2024release} have demonstrated remarkable abilities in tasks  ranging from drafting emails to writing code, showcasing the versatility of GenAI in understanding and generating natural language.
\citep{noy2023experimental} reported that ChatGPT can substantially raise productivity by decreasing the average time consuming on mid-level professional writing tasks by 40\% and increasing the output quality by 18\%.

Apart from natural language generation, GenAI models have been used for generating realistic images. 
For instance, GANs have been used to create high-resolution, photorealistic images that are indistinguishable from real photographs \citep{karras2019style}. 
Such capabilities have enabled applications in fields like art generation \citep{louie2020novice}, virtual reality \citep{hashim2023revolutionizing}, and data augmentation used for other AI model trainings \citep{gowda2024data}.

As with many other domains, the capabilities of GenAI have recently been explored in the field of supply chain management, albeit very few studies exist. Those which does exist, do not yet report technical performance, but rather explore potential benefits and applications.
\citep{wamba2023both} explored the benefits, challenges and trends associated with GenAI technologies like ChatGPT in Supply Chain and Operation Management (SCOM) by surveying practitioners in the United Kingdom and the United States.
This study reveals an increased efficiency from GenAI adopters compared to non-adopters and highlights that the integration of GenAI can significantly enhance overall supply chain performance.
A subsequent study \citep{fosso2024chatgpt} extended the exploration by additionally mapping the maturity levels of GenAI projects across supply chains and identified the specific operational benefits and challenges that organizations need to overcome.
Meanwhile, \citep{jackson2024generative} provided a comprehensive understanding of both AI and GenAI functionalities and applications in the SCOM context.
It also offers a practical framework for both practitioners and researchers to identify where and how AI and GenAI can be applied in SCOM to enhance decision-making processes, optimize operations, prioritize investments, and develop necessary skills.

In the industry, companies early adopters of GenAI have reported enhancement of their supply chain task performance. 
Mars applied an GenAI platform offered by Celonis \citep{celonis2024genai} to optimize truck loads and reduce manual efforts by 80\% and improve delivery efficiency \citep{mars2023genai}.
Amazon leveraged GenAI to streamline and improve the delivery process \citep{meiyappan2021position} while FedEx applied GenAI to generate more precise package arrival estimates \citep{fedeX2023genai}. 
Shein in the fast fashion sector leveraged GenAI to understand the changes in customer demand and interest, allowing it to adjust its supply chain in real time \citep{shein2023genai}.
A report from McKinsey \citep{mckinsey2023genai} shows that GenAI could add up to \$275 billion to the operating profits of apparel, fashion, and luxury sectors in the next three to five years.

Inspired by such achievements of GenAI in both academic and industry, we explore how GenAI models can enhance supply chain visibility which is a crucial challenge in supply chains \citep{pichler2023building}.

Language models as a type of GenAI models have been pointed out the great potential of improving supply chain performance \citep{srivastava2024exploring, aguero2024potential}. 
One of the earlier applications in the supply chain domain has been supply chain mapping wherein \citep{wichmann2018towards} automatically extracted structured supply chain information from unstructured natural text to answer questions such as ``who supplies whom with what from where?'', indicating that language models can help extract relationships of entities in supply chain networks.

Various researchers showed that relational knowledge that has been learnt in  pretrained language models (pretrained LMs) can allow the recovery of factual knowledge from them \citep{petroni2019language, bouraoui2020inducing, safavi2021relational}. 
This points to the potential opportunity to retrieve relational knowledge of entities in supply chain networks from pretrained LMs to predict hidden dependencies.

As pretrained LMs are trained on massive data from diverse sources, the learned knowledge in these models is general and not capable of a task that requires specialised domain knowledge. 
To solve this problem, two solutions can be used. 
One is training a language model for a specific task and the other one is fine-tuning a pretrained LM for a specific task. 
The former requires large computational resources and data, and is also time-consuming, prompting researchers to advocate the latter option of fine-tuning \citep{fichtel2021prompt, yasunaga2022linkbert}. 
Relevant examples include \citep{yasunaga2022linkbert}, who fined-tuned a BERT model to predict links among documents and \citep{tan2024walklm} developed a uniform framework to fine-tune language models on multiple tasks including link prediction. 
Results from both works showed that fine-tuning pretrained LMs can accurately predict links. However, fine-tuning pretrained LMs requires expensive expertise in NLP, and most companies in supply chains are small and midsize enterprises (SMEs) limited by the budget to employ such expensive experts. 
Additionally, SMEs typically lack access to sufficient high-quality labelled data for effective fine-tuning and are constrained by limited computational sources.

In addition, another concern for using pretrained LMs is the issue of ``hallucination'' leading to the generated outputs being not factual or inaccurate \citep{huang2023survey}. 
The phenomena of ``hallucination'' results from the biases and inaccuracies in massive amounts of training data from various sources. 
In the supply chain context, such hallucinations can lead to erroneous demand forecasts or misinterpretation of supply chain relationships, potentially resulting in operational disruptions and financial losses. 
For instance, a generative model might incorrectly predict a surge in demand based on fabricated trends, leading to overproduction and increased inventory costs.

\citep{agrawal2023can, martino2023knowledge, guan2024mitigating} found that knowledge graphs that organise information into entities and relationships, creating a network of interconnected facts, can help mitigate the hallucination issue by providing a structured, verifiable source of information that the language models can reference during text generation. Contextual information represented by linkages in knowledge graphs can enhance the language models' understanding of the relationships of different concepts to ensure the generated outputs being consistent with the established facts within knowledge graphs.

Building on prior research works that highlight the potential of relational knowledge learned in pretrained LMs  \citep{petroni2019language, bouraoui2020inducing, safavi2021relational} and addressing the hallucination issue that knowledge graphs can alleviate, we develop a new approach that combines pretrained LMs with traditional machine learning techniques to predict multiple interconnected relationships within supply chain networks represented by knowledge graphs. 
This new approach does not require fine-tuning. 
The pre-trained language models are used to convert textual data into high-dimensional vector embeddings that capture semantic meanings and contextual nuances. 
These embeddings are rich in linguistic information but may lack domain-specific factual accuracy when used in isolation, leading to hallucinations.
The machine learning models learn to map the semantic embeddings to the factual relationships represented in the knowledge graph. 
This integration ensures that the predictions are not solely based on language patterns but also aligned with the actual data from the knowledge graph. 
The knowledge graph acts as a factual anchor, constraining the model to produce outputs consistent with known supply chain relationships.
It can reduce the risk of hallucinations by providing a factual basis for relationship predictions, enhancing the reliability and accuracy of the model, and also leverage the strengths of pre-trained language models in understanding and encoding contextual information while counteracting their tendency to generate incorrect information when used alone.

\subsection{Link Prediction in Supply Networks} \label{sec: link prediction}

Supply chains are complex networks that exhibit non-linear interactions and inter-dependencies among various entities, processes and resources  \citep{choi2001supply}. 
The large-scale and non-linear nature of these networks often hinder their visibility, which, in return, makes the identification of potential risks challenging. 
Researchers have shown that predicting relationships of entities within a supply chain network can contribute to improved visibility \citep{brintrup2018predicting}.

Formally, researchers framed the problem of identifying relationships in a supply chain as a \textit{link prediction} problem on a graph. 
Link prediction is widely used to solve problems in various domains such as social networks, where it predicts potential connections between individuals based on existing ties, interests, or behaviours \citep{hasan2011survey}; recommendation systems, where it identifies associations among product sales \citep{su2020link}; and biological networks, where it predicts interactions among genes, proteins, and other biological entities \citep{cocskun2021node}.
The approaches used in the majority of these works are similarity-based and learning-based. 
Similarity-based approaches predict the connection of two nodes by leveraging the similarity of characteristics between two nodes \citep{zareie2020similarity}, while learning-based approaches involve training machine learning models to infer the likelihood of connections between nodes based on node and network level features \citep{ahmed2016supervised}.  Supply chain researchers have argued that similarity-based approaches are unsuitable for predicting links in this domain as similar companies usually do not connect due to competition, advocating for learning-based approaches.

\citep{brintrup2018predicting} used two machine learning models to develop a link prediction approach to predict supplier interdependencies in a manufacturing supply network.
\citep{kosasih2022machine} developed a machine learning model using Graph Neural Network (GNN) to detect potential links that are unknown to the buyer in an automotive supply chain network. 
Later on, \citep{kosasih2022towards} proposed a neurosymbolic machine learning method using a combination of GNN and knowledge graph reasoning to predict multiple types of links in two supply chain networks. 
In the same year, \citep{brockmann2022supply} also performed link prediction using GNN but on an uncertain supply chain knowledge graph. 
\citep{brintrup2018predicting} and \citep{kosasih2022machine} predicted one type of relationship. \citep{kosasih2022towards} and \citep{brockmann2022supply} predicted multiple types of relationships on knowledge graphs.  
Furthermore, \citep{mungo2023reconstructing} considered production network reconstruction as link prediction and used Gradient Boosting to predict hidden relationships to reconstruct the production network. 
Another work that focuses on the firm-level network reconstruction is from \citep{ialongo2022reconstructing}, in which a generalised maximum-entropy reconstruction method is introduced to reconstruct the firm-level network based on partial information.

While these approaches have been very valuable, the mere identification of a transactional relationship between companies does not allow sufficient contextual information for actionable insights. 
Considering the case that Toyota and Hewlett Packard (HP) are predicted to share a transactional relationship, it might be that HP has sold a large number of office printing equipment to Toyota and it might also be that Toyota uses HP's 3D printing material in its production. 
For an analyst looking to identify whether a disruption at HP would cause an issue to the automotive industry, the two types of relationships would have different implications on actual production output. Similar issues arise when we consider other contextual information such as production locations. In this work, we will focus on adding context to identified relationships by a type of GenAI models, pretrained LMs, as a knowledge base \citep{srivastava2024exploring}.

Our hypothesis is that doing so will allow us to combine the structured factual knowledge that can be obtained from an ontological presentations afforded by knowledge graphs, with the general unstructured knowledge base that can be obtained from pretrained LM, thereby mitigating the hallucination issue caused by using pretrained LMs alone. 
Besides, it will allow us to enhance the supply chain visibility by a complete supply chain knowledge graph for a comprehensive understanding of the supply chain dynamics.



\section{Combining Pretrained Language Models and Knowledge Graphs} \label{sec: methodology}


\subsection{Preliminaries: From triplets to quintuplets} \label{sec: problem definition}

\begin{figure*}[th!]
    \centering
    \includegraphics[width=0.72\textwidth]{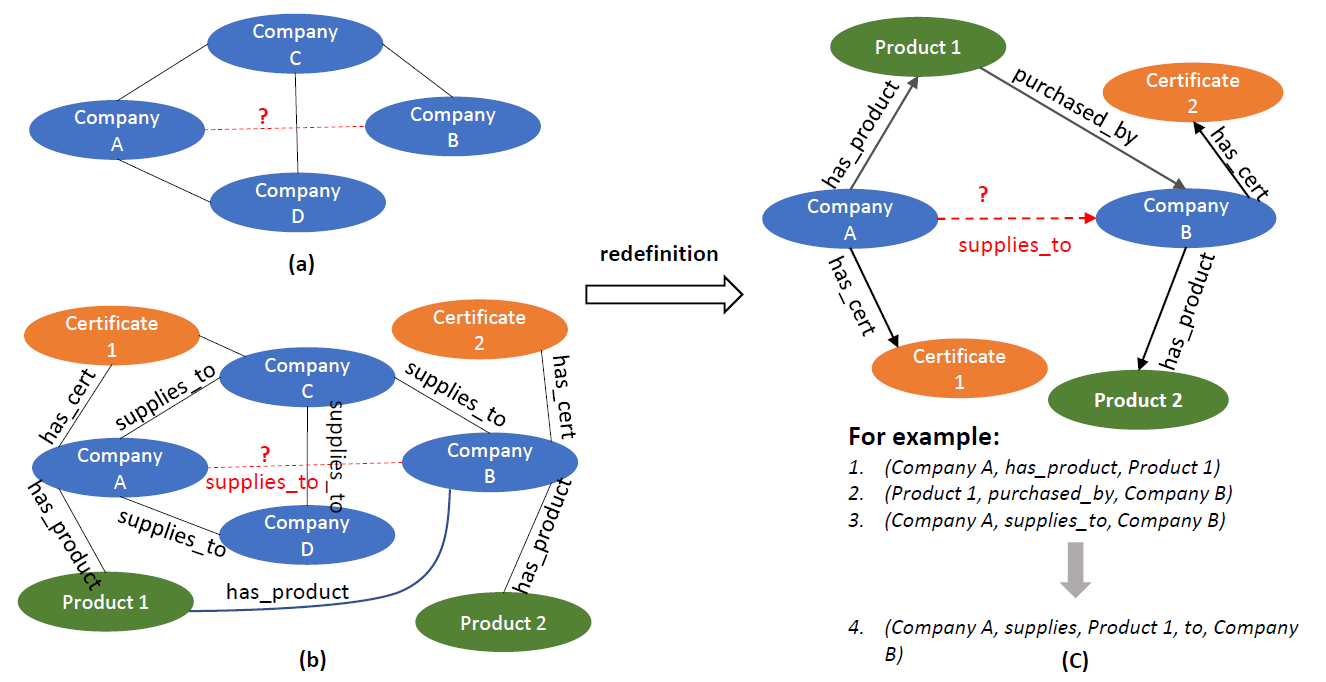}
    \caption{(a) company-level relationships in a supply chain network~\citep{brintrup2018predicting}; (b) multiple relationships including (company, supplies to, company), (company, has, certificate) and (company, has, product) in a knowledge graph~\citep{kosasih2022towards}; (c) quintuplet based relationships on a knowledge graph. For example, company A supplies product 1 to company B, and company A with certificate 1 has product 1. \newline 
    Figure 1 Alt-text: The diagram shows company-level relationships within a supply chain network, multiple relationships in a supply chain knowledge graph, and quintuplet-based relational structures within the knowledge graph.}
    \label{fig: relationship}
\end{figure*}

As mentioned earlier, the first set of studies in literature solved the supply chain link prediction problem on a graph with links representing companies \citep{brintrup2018predicting, kosasih2022machine, cai2021line, kazemi2018simple}. 
This approach aimed at learning connection patterns surrounding two companies, estimating a connection between them (Figure~\ref{fig: relationship}). 
The second set of studies represent supply chain information as a knowledge graph (KG) with multiple types of links \citep{rossi2021knowledge, kosasih2022towards}.

A KG is represented by a heterogeneous graph G and its ontology O, the former being the actual data and the latter its schema. 
KG can also be seen as a collection of facts represented by predicate logic statements. 
A KG is based on an ontology, that defines data types and attributes, with a relational taxonomy. 
Each item in the data is an entity (or a node in a graph), and the relationships between data are edges, or links. 
In previous works, KGs have been used to model edges such as who-produces-what, who-has-what-certification, in addition to buyer-supplier links (Figure~\ref{fig: relationship}), the structure of which then inform one another.

Both of the above approaches introduce triplets to describe and predict relationships. 
A triplet, also known as a triple or a statement, consists of three components: subject, predicate, and object, and is used to define the relationship between a subject and an object (Figure~\ref{fig: relationship}). 
Singular links are analysed at a time - where we can predict \textit{``what a company produces''}, and \textit{``to whom it sells''}, but not contextual information such as \textit{``which product a company sells to its buyer in which location''}. 
Understanding context in supply chains is important to accurately predict risks.

To define context, we propose a new term, ``quintuplet'', where information that is represented by three triplets, \textit{(Company A, has\_product, Product 1), (Product 1, purchased\_by, Company B)} and \textit{(Company A, supplies\_to, Company B)}, can be condensed into \textit{(Company A, supplies, Product 1, to, Company B)}.

Given a knowledge graph $G(\mathcal{V}, \mathcal{E})$, $\mathcal{V}$ is the set of entities and $\mathcal{E} \subseteq (\mathcal{V} \times \mathcal{V})$ is a set of relationships. 
The relationship between entity $v_1$ and entity $v_2$, in a knowledge graph can be represented by a triplet $(v_1, \epsilon_{1,2}, v_2)$, in which $v_1, v_2 \subseteq \mathcal{V}$ and $\epsilon_{1,2} \subseteq \mathcal{E}$.

In contrast, a quintuplet would inform, $(v_1, \epsilon_{1,2}, v_2, \epsilon_{2,3}, v_3); \\ v_1, v_2, v_3 \subseteq \mathcal{V}$ and $\epsilon_{1,2}, \epsilon_{2,3} \subseteq \mathcal{E}$, which becomes the target of the prediction.  
Compared to a triplet representing one relationship, a quintuplet describes multiple connected relationships. 
The multiple connected relationships and the connected entities in a quintuplet can represent a small subgraph of the knowledge graph, leading to contextual information for an entity or relationship.

\subsection{The pretrained LM-based Machine Learning Framework}

\begin{figure*}[th!]
    \centering
    \includegraphics[width=1.0\textwidth]{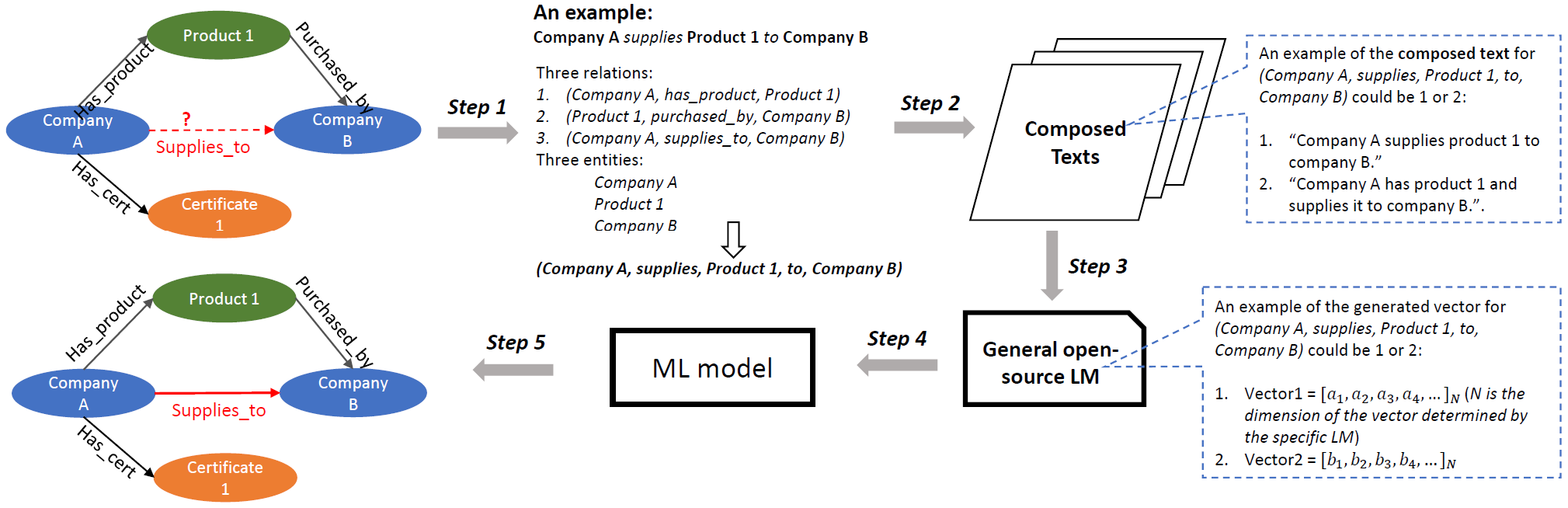}
    \caption{The pretrained LM-enhanced supply chain link prediction framework. \newline Figure 2 Alt-text: The diagram presents the pretrained LM-enhanced supply chain link prediction framework.}
    \label{fig: fremawork}
\end{figure*}

We propose a pretrained LM-based machine learning framework which transfers a knowledge graph described by triplets into quintuplets to generate composed texts, and then sends these composed snippets of text to a pretrained LM to retrieve the relational knowledge that has been learned a priori (Figure~\ref{fig: fremawork}). 
The retrieved relational knowledge is represented by vectors of fixed length, that are further learned by a machine learning model to predict the multiple connected relationships of entities represented by quintuplets in supply chain networks.

We begin by constructing a knowledge graph from already known data that characterises the supply chain. 
This may involve but is not limited to a priori known supply-buy relationships, products and certifications, and to a large extent determined by data that is available to the researcher. 
The original data used to construct the supply chain knowledge graphs are commonly collected from various sources such as Enterprise Resource Planning (ERP) systems, transaction records, market reports, and social media (see \cite{brintrup2023digital} for a review). 
The supply chain data used in this work was collected by a third-party data provider. 
To construct a supply chain knowledge graph, we need to define the ontology using the data that can character supply chain information. 
In this case, the ontology includes the definitions of entities, i.e. \textit{companies}, \textit{products} and \textit{certifications}, and the relationships between these entities, such as \textit{has\_product}, \textit{purchased\_by}, \textit{has\_cert}, and \textit{supplies\_to}, shown in Figure~\ref{fig: fremawork}. 
The knowledge defined by the ontology is structured into triplets, and each triplet is an instance of the relationships and entities defined by the ontology.

We define the quintuplets to represent the contextual knowledge that we aim to predict. 
As a case example, we use product flow on supply-buy links, however a quintuplet can also represent contextual information on locations, types and volumes of transactions, depending on the question at hand.

We then reconstruct relationships originally represented by triplets, into quintuplets. 
For example three triplets \textit{(Company A, has\_product, Product 1), (Product 1, purchased\_by, Company B)} and \textit{(Company A, supplies\_to, Company B)} can be used to generate a quintuplet of the sort: \textit{(Company A, supplies, Product 1, to, Company B)}. 
Another example is: \textit{(Company A, with, Certificate 1, has, Product 1)} generated by two triplets, \textit{(Company A, has\_cert, Certificate 1)} and \textit{(Company A, has\_product, Product 1)}.

The next step involves transferring quintuplets into composed snippets of text with a user-defined schema. 
For example, a quintuplet of \textit{(Company A, supplies, Product 1, to, Company B)} can be transferred into \textit{Company A supplies product 1 to company B} or \textit{Company A has product 1 and supplies it to company B}. 
The text used to represent a quintuplet can include different types of sentences but needs to be contextually accurate and consistent.

The composed text is then sent to a pre-trained language model for embedding and  and retrieved hidden relational knowledge previously learned in the pre-trained language model. 

The embeddings of quintuplets with retrieved relational knowledge are used to train a suitable machine learning model for quintuplet prediction. 
As pretrained LMs cannot directly predict factual relationships in a supply chain, we use a machine learning model for it.
The resulting trained model can then be used for quintuplet prediction.

\subsubsection{Language Model Selection}\label{subsec: LM selection}

We select five general open-source pre-trained language models for experimentation with the following considerations:

\begin{enumerate}

    \item[1)] \textit{Diversity}: We test several pretrained LMs to investigate whether our approach is applicable across the state of the art.
    
    \item[2)] \textit{Model size}: Many existing works in the field of NLP \citep{shoeybi2019megatron, shin2020biomegatron, narayanan2021efficient} suggest that larger model size can lead to improved performance. Therefore, pretrained LMs with different model sizes are selected for experimentation. 

    \item[3)] \textit{Output dimensions}: Increasing the dimension of a language model can potentially capture more complex patterns and nuances in the data, however, cannot guarantee better representation by default \citep{kenton2019bert}. 
    Thus, selected pretrained LMs have different dimensions of their representations for evaluation.   

\end{enumerate}

Based on the considerations above, we select five pretrained LMs that all were developed on the basis of Transformer \citep{vaswani2017attention} but have different model sizes and output dimensions (Table~\ref{tab: language models}). 

\begin{table*}[th!]
\centering
\captionof{table}{Selected pretrained language models}
\small
    \begin{tabular}{|p{2.78cm}|p{0.8cm}|p{2.40cm}|p{1.4cm}|p{1.3cm}|p{3.5cm}|} 
    \hline
     Model Name & Model Size &  Dimensions & Max Sequence Length & Training Data Size & Reference \\
    \hline
    distiluse-base-multilingual-cased-v2 & 480MB & 1 million sentence pairs (15 languages) & 512 & 128 & \citep{reimers-2019-sentence-bert} \\
    all-distilroberta-v1 & 290MB & over 1 billion pairs & 768 & 512 &  \citep{liu2019roberta}\\
    all-MiniLM-L12-v2 & 120MB &over 1 billion pairs & 384 & 256 &  \citep{wang2020minilm} \\
    all-MiniLM-L6-v2 & 80MB &  over 1 billion pairs  & 384 & 256 & \citep{wang2020minilm} \\
    paraphrase-albert-small-v2 & 43MB & 16GB of uncompressed text & 768 & 256 & \citep{lan2019albert}\\
     \hline
    \end{tabular}
\label{tab: language models}
\end{table*}

Among the selected pretrained LMs, ``paraphrase-albert-small-v2'' is the smallest with 43 MB and a six-layer version of ``albert-base-v2'' that originates from \citep{lan2019albert} aiming to solve the problems of GPU/TPU memory limitations and longer training times by lowering model size. 
Compared to the original BERT, ``albert-base-v2'' introduces two parameter reduction techniques to reduce memory consumption and increase training speed.

The second smallest model is ``all-MiniLM-L6-v2'', a six-layer version of \citep{wang2020minilm}, developed by compressing the large Transformer-based pre-trained model using a simple but effective approach called deep self-attention distillation \citep{wang2020minilm}. 
It introduces the conceptions of the student model and the teacher model. 
``all-MiniLM-L6-v2'' referred to as the student model in \citep{wang2020minilm} is trained by mimicking the self-attention module, in Transformer networks, of the large language model referred to as the teacher model and also by distilling the self-attention module of the last Transformer layer of the teacher model. 
In addition, ``all-MiniLM-L6-v2'' only keeps 50\% of parameters of the teacher model but can retain more than 99\% of accuracy on several benchmark tasks \citep{wang2020minilm}. 
``all-MiniLM-L12-v2'' is similar to ``all-MiniLM-L6-v2'' but is a 12-layer version of \citep{wang2020minilm}, leading to big model size with more parameters.

``all-distilroberta-v1'' is a distilled version of the BERT base model in \citep{kenton2019bert}. 
It is smaller and faster than BERT but developed using the BERT base model as a teacher. 
Compared to ``paraphrase-albert-small-v2'', ``all-MiniLM-L6-v2'' and ``all-MiniLM-L12-v2'', this model size is larger but smaller than ``distiluse-base-multilingual-cased-v2''. 
``distiluse-base-multilingual-cased-v2'' is also a modification of the pretrained BERT network, but trained on the data in 15 languages \citep{reimers-2019-sentence-bert}, compared to ``all-distilroberta-v1'' being trained in English.

\subsubsection{Machine Learning Model Selection} \label{subsec: machine learning selection}

Five machine learning models are selected based on the suitability of the model for application to link prediction and past works. 
These include Artificial Neural Network (ANN) \citep{yegnanarayana2009artificial}, Convolutional Neural Network (CNN) \citep{albawi2017understanding}, Logistic Regression (LogReg) \citep{kleinbaum2002logistic}, Long Short-Term Memory (LSTM) \citep{hochreiter1997long}, and AutoEncoder \citep{wang2014generalized}. 
Their architectures and parameter settings are presented in Table~\ref{tab: five models}.

ANN is composed of three linear layers with 300 neurons each and each linear layer is followed by a Batch Normalization layer (BatchNorm) and a Rectified Linear Unit (``ReLU'') activation function. 
BatchNorm aims to normalise the output of each linear layer to ensure a more stable training process while the function ``ReLU'' contributes to the acceleration of the training phase and the mitigation of the problem of vanishing gradients. 
The output from the last ``ReLU'' activation function is then sent to a fully-connected (FC) layer before being coupled with a ``Softmax'' function to achieve the final prediction.

The CNN model consists of three convolutional layers, each followed by the ``ReLU'' function, an Average Pooling layer (AvgPooling), and a BatchNorm layer. 
``ReLU'' and ``BatchNorm'' function the same as them in ANN while AvgPooling layer serves to decrease the dimensionality of the features outputted by the ``ReLU''. 
Similar to ANN, a FC layer followed by a Softmax function is used to output the final prediction. 
Considering the computation cost and the performance, the first Conv1D layer is configured with 32 kernels of size 7 and of stride 2 while the second and third Conv1D layers use 64 kernels of size 7 and of stride 1.

The LogReg model has two linear layers, followed by a logistic function ``Sigmoid''. 
The first linear layer with 200 neurons and the second with 2 neurons are used to analyse the relationship between output and input features. 
The LSTM model contains two LSTM layers with bi-directions, followed by the FC layer. 
The input and hidden sizes in each layer are set as 16. The AutoEncoder model consists of an encoder for transforming the input to a compressed representation, a decoder for reconstructing the original input from the encoded representation, and an FC layer followed by the Softmax function for the final link prediction. 
Both encoder and decoder are composed of two linear layers, each followed by the ``ReLU'' function. 
To encode input to a compressed representation, the number of neurons in two linear layers are respectively 96 and 48. 
In the decoder, two linear layers with 48 and 96 respectively are used to reconstruct the original input.

Since our task considers link prediction as a binary classification problem, all models excluding LogReg use a FC layer with 2 neurons.

\begin{table*}[ht!]
\caption{The architectures and parameters of five selected common machine learning models.} 
\label{tab: five models}
\centering
\begin{tabular}{|p{1.7cm}|p{0.9cm}|p{1.7cm}|p{1.0cm}|p{1.2cm}|p{0.9cm}|p{1.2cm}|p{1.0cm}|p{2.6cm}|p{0.9cm}|}
\hline
\multicolumn{2}{|c}{\textbf{ANN}} & \multicolumn{2}{|c}{\textbf{CNN1D}} & \multicolumn{2}{|c}{\textbf{LogReg}} & \multicolumn{2}{|c|}{\textbf{LSTM}} & \multicolumn{2}{|c|}{\textbf{AutoEncoder}} \\
\hline
Layer Name & \makecell{Param\\-eters} & Layer Name & \makecell{Param\\-eters} & Layer Name & \makecell{Param\\-eters} & Layer Name & \makecell{Param\\-eters} & Layer Name & \makecell{Param\\-eters}\\

\hline
Linear\textsuperscript{1}    & 300 & Conv1D\textsuperscript{1}     & 32,(7),2 & Linear\textsuperscript{1} & 200 & LSTM\textsuperscript{1} & 16,16,bi & Linear\textsuperscript{1} (Encoder) & 96 \\
BatchNorm\textsuperscript{1} & 300 & ReLU\textsuperscript{1}       & --       & --                        & --  & --                      & --       & ReLU\textsuperscript{1} (Encoder)   & -- \\
ReLU\textsuperscript{1}      & --  & AvgPooling\textsuperscript{1} & -,(7),2  & --                        & --  & --                      & --       & Linear\textsuperscript{2} (Encoder) & 48 \\
--                           & --  & BatchNorm\textsuperscript{1}  & 32       & --                        & --  & --                      & --       & ReLU\textsuperscript{2} (Encoder)   & -- \\

\hline
Linear\textsuperscript{2}    & 300 & Conv1D\textsuperscript{2}     & 64,(7),1 & Linear\textsuperscript{2} & 2   & LSTM\textsuperscript{2} & 16,16,bi & Linear\textsuperscript{1} (Decoder) & 48 \\
BatchNorm\textsuperscript{2} & 300 & ReLU\textsuperscript{2}       & --       & --                        & --  & --                      & --       & ReLU\textsuperscript{1} (Decoder)   & -- \\
ReLU\textsuperscript{2}      & --  & AvgPooling\textsuperscript{2} & -,(7),1  & --                        & --  & --                      & --       & Linear\textsuperscript{2} (Decoder) & 96 \\
--                           & --  & BatchNorm\textsuperscript{2}  & 64       & --                        & --  & --                      & --       & ReLU\textsuperscript{2} (Decoder)   & -- \\

\hline
Linear\textsuperscript{3}    & 300 & CNN-1D\textsuperscript{3}     & 64,(7),1 & --                        & --  & --                      & --       & --                                  & -- \\
BatchNorm\textsuperscript{3} & 300 & ReLU\textsuperscript{3}       & --       & --                        & --  & --                      & --       & --                                  & -- \\
ReLU\textsuperscript{3}      & --  & AvgPooling\textsuperscript{3} & -,(7),1  & --                        & --  & --                      & --       & --                                  & -- \\
--                           & --  & BatchNorm\textsuperscript{3}  & 64       & --                        & --  & --                      & --       & --                                  & -- \\

\hline
--                           & --  & Flatten                       & --       & --                        & --  & --                      & --       & --                                  & -- \\

\hline
FC                           & 2   & FC                            & 2        & --                        & --  & FC                      & 2        & FC                                  & 2  \\ 
Softmax                      & --  & Softmax                       & --       & Sigmoid                   & --  & Softmax                 & --       & Softmax                             & -- \\

\hline
\end{tabular}
\end{table*}

\section{Case Study}\label{sec: case study}

The case study used to evaluate the proposed approach from the automotive sector where companies produce car parts, such as engines, front axles, fuel tanks and sell them to car manufacturing companies around the world. 
The dataset has been used as a benchmark for link prediction problem within supply chain networks in previous works and therefore offers potential for cross-comparison \citep{brintrup2018predicting, kosasih2022machine, kosasih2022towards}. 
The dataset comprises 43,131 companies spanning 72 countries and producing 927 distinct products, each company associated with one or more of 5 certification types, along with the relationships among these entities (Table~\ref{tab: marklines data}). 

\begin{table}[th!]
\centering
\captionof{table}{Basic descriptions of Marklines data}
\small
    \begin{tabular}{|p{1.2cm}|p{4.5cm}|p{1.2cm}|} 
    \hline
     \textbf{Entity} & \textbf{Example} & \textbf{Unique Number} \\
    \hline
    Company & Hamenz For German Tech. Ind. (S.A.E.) & 43,131 \\
    Country & Egypt & 79 \\
    Certificate & IS09001, QS9000, ISO/TS16949 & 5 \\
    Product & Piston Ring Machining & 927 \\
     \hline
    \end{tabular}
\label{tab: marklines data}
\end{table}

We separated the data at the country level so as to evaluate our approach over multiple heterogeneous datasets. 
As shown in Table~\ref{tab: detailed data} each partition has different numbers of companies, products, certificates and relationships.

\begin{table}[!ht] 
\caption{Data description for each country.}
\centering
\begin{filecontents*}{tables/data_description.csv}
,company,product,certification,numberOfRelations,country
13,150,251,4,1997,AUSTRALIA
23,195,196,5,3199,AUSTRIA
24,293,245,5,4704,BELGIUM
0,603,458,5,10705,BRAZIL
12,390,359,5,22734,CANADA
9,10500,877,5,330180,CHINA
11,329,281,5,3395,CZECH REPUBLIC
7,518,510,5,21052,FRANCE
25,1965,769,5,114793,GERMANY
21,363,283,5,3820,HUNGARY
15,1509,646,5,61539,INDIA
6,486,380,4,12020,INDONESIA
18,389,425,5,15264,ITALY
10,5068,890,5,215514,JAPAN
16,1195,706,5,38295,KOREA
5,384,355,4,5481,MALAYSIA
17,463,395,5,7900,MEXICO
20,456,303,5,3467,POLAND
22,213,235,4,2038,RUSSIA
4,185,242,5,2623,SOUTH AFRICA
2,282,385,5,12044,SPAIN
26,236,224,5,6802,SWEDEN
3,777,488,5,15982,TAIWAN
14,1296,535,4,20439,THAILAND
19,449,369,5,13141,TURKEY
8,2577,800,5,91292,U.S.A.
1,932,502,5,15836,UK
\end{filecontents*}
\csvreader[%
before reading=\footnotesize\setlength{\tabcolsep}{1.5pt},
 tabular={|p{2.5cm}|p{1.4cm}|p{1.4cm}|p{1.4cm}|p{1.4cm}|},
        table head = \hline\textbf{Country Name} & \textbf{{Num-Company}} & \textbf{{Num-Product}} & \textbf{Num-Certificate} & \textbf{Num-Relations}\\\hline,
        late after line= \\,
        late after last line=\\\hline %
        ]{tables/data_description.csv}{country=\country, company=\company, product=\product,certification=\certification, numberOfRelations=\numberOfRelations}%
        {\country & \company & \product & \certification & \numberOfRelations}
       \centering
    \label{tab: detailed data}
\end{table}

27 datasets have thus been generated. 
As a starting point, we use the same knowledge graph ontology as previous works with three types of entities: companies, products, and certificates, and four types of relationships (triplets): \textit{(company, has\_product, product)}, \textit{(company, has\_cert, certificate)}, \textit{(company, supplies\_to, company)} and \textit{(product, purchased\_by, company)} (see Figure~\ref{fig: relationship} c). 

Four triplets are used to generate two quintuplets for the evaluation of the proposed approach. 
Two quintuplets are \textit{(company, supplies, product, to, company)} and \textit{(company, with, certificate, has, product)}. 
The prediction problem thus is the existence of a given quintuplet. 
Therefore, we consider the relationship prediction in a quintuplet as a binary classification problem.      

Next, we explain how we generate positive and negative relationships based on quintuplets to train the models, followed by experimental settings and results.

\subsection{Generating training data}\label{sub: data description}

We refer to actual relationships in a quintuplet as positive relationships and non-existing relationships as negative relationships. 
To train a machine learning model both positive relationships and negative relationships are needed.

Negative quintuplets are generated by the same triplets that were used to produce positive quintuplets. 
Consider the following positive quintuplet, \textit{(company A, supplies, product 1, to, company B)}, three negative quintuplets can be generated by replacing any one of the three entities: \textit{(company A, has\_product, product 1)}, \textit{(product 1, purchased\_by, company B)} and \textit{(company A, supplies\_to, company B)}, using the one that does not connect the other two. 

Given two lists of unique entities i.e. the list of unique companies and the list of unique products, \textit{(company A, supplies, product 1, to, company B)}, we randomly select a company named company C, that does not connect both company A and company B, from the list of unique companies to replace company A or company B. 
Alternatively, we can randomly select a product named product 2, that does not connect to either company A and company B, and replace it with product 1. 
Therefore, we can generate three negative quintuplets: \textit{(company C, supplies, product 1, to, company B)}, \textit{(company A, supplies, product 1, to, company C)} and \textit{(company A, supplies, product 2, to, company B)}. 
In addition, incorrect relationship direction is also considered as negative quintuplets, such as \textit{(company B, supplies, product 1, to, company A)}.

Based on the above method each positive quintuplet can be used to produce several negative quintuplets. 
Since negative training data would far outweigh positive training data, we randomly select one negative quintuplet with the related positive quintuplet to train the model in order to have a balanced dataset.      

\subsection{Experimental Settings} \label{sub: experimental settings}

Experimental settings in this work include benchmark settings and model training settings. 
The benchmark settings aim to evaluate whether machine learning models with the help of pretrained LMs can provide more accurate relationship predictions in supply chain networks while the model training settings aim to set the optimal parameters during the model training phase.

\subsubsection{Benchmarks}
To evaluate the effectiveness of our proposed approach, we set machine learning models without the help of pretrained LMs as benchmarks. 
The proposed approach is designed to power machine learning models by pretrained LMs so we also select five pretrained LMs (cf. Section~\ref{sec: methodology}) to test the approach. 

\subsubsection{Settings of Model Training}

As the language models used in this work are pre-trained models, we only need to set parameters to train machine learning models, which are shown in Table~\ref{tab: five models}. 
The parameters during the training phase include the number of epochs, $E$, batch size, $B$, and learning rate, $r$. 
To ensure the uniformity of experiments and follow common guidelines in machine learning training \citep{yang2020hyperparameter}, we set $B$ as 64 and $r$ as 0.001 for all five machine learning models. 
For the number of epochs, $E$, we use the $stop-early$ strategy to stop the training process if the training loss decreases but validating loss increases in 10 continuous epochs for the determination of $E$ and also prevents overfitting.

We consider the problem of relationship prediction in supply chain networks as a binary classification problem mentioned earlier. 
We thus use the \textit{Cross-Entropy Loss} as the loss function for all machine learning model training. 
\textit{Adam} \citep{kingma2014adam} is selected as the optimiser for all machine learning models.

In addition, following the common rules for splitting dataset into training, validating and testing, we use 70\%, 10\% and 20\% of relationships present in each data partition. 
All experiments are run on a desktop with an ${Intel}^{R}~{Core}^{TM}$ i9-9900K CPU and a GeForce RTX 2080 Ti GPU card with 11 GB physical memory. 
PyTorch is used to develop and train all models.

\subsubsection{Performance Metrics}
\label{subsec: performance metrics}

Common metrics, including \textit{accuracy}, \textit{precision}, \textit{recall} and \textit{f-score}, used to evaluate the performance of a classification approach are also used to evaluate our proposed approach. 
Table~\ref{tab: confusion matrix} shows a confusion matrix used to calculate the four metrics. 
In our case, \textit{TP} and \textit{TN} respectively represent positive and negative relationships that are predicted correctly while \textit{FP} and \textit{FN} respectively describe positive and negative relationships that are predicted incorrectly. 
Based on Table~\ref{tab: confusion matrix}, the four common metrics can be calculated as:

\begin{itemize}
    \item \textit{accuracy} $= \frac{\textit{TP} + \textit{TN}}{\textit{TP} + \textit{TN} + \textit{FP} + \textit{FN}}$ describes the ratio of correct relationship predictions to the total number of relationships.
    
    \item \textit{precision} $= \frac{\textit{TP}}{\textit{TP} + \textit{FP}}$ stands for the ratio of accurate predictions of positive relationships to the total number of predicted positive relationships.  
    
    \item \textit{recall} $= \frac{\textit{TP}}{\textit{TP} + \textit{FN}}$ represents the ratio of accurate predictions of positive relationships to the total number of positive relationships.
    
    \item \textit{f-score} shows the equilibrium between the precision and the recall, $\frac{2 \times \textit{Precision} \times \textit{Recall}}{\textit{Precision} + \textit{Recall}}$. 
\end{itemize}

\begin{table}[th!]
\captionof{table}{Confusion Matrix}
\label{tab: confusion matrix}
\begin{tabular}{|l|c|c|}
\hline
 & Predicted Positive & Predicted Negative\\
\hline
Real Positive & \textit{True Positive} (\textit{TP}) & \textit{False Negative} (\textit{FN})  \\
\hline
Real Negative & \textit{False Positive} (\textit{FP}) & \textit{True Negative} (\textit{TN}) \\
\hline
\end{tabular}
\end{table}

When we split the dataset into training, validating and testing, we randomly shuffle all positive and negative relationships for fairness. 
This process may lead to an imbalanced testing dataset even though the overall dataset is balanced. 
Therefore, to truly reflect the performance of our approach, we use \textit{weighted~f-score} shown in Equation~(\ref{eq: fw-score}) to replace the commonly used \textit{f-score}.

\begin{equation} \label{eq: fw-score}
    f_w-score = \sum_k 2 \times w_k \frac{{Precision}_k \times {Recall}_k}{{Precision}_k + {Recall}_k}
\end{equation}

\noindent where $w_k$ is the ratio of relationships for class $k$ over all relationships and is equal to $\frac{n_k}{N}$. $N$ is the total number of relationships while $n_k$ is the number of relationships in class $k$.

In addition, we expect the developed approach to equally consider the importance of positive and negative relationships. 
Thus, we follow \citep{grandini2020metrics}, who compared metrics for multi-class classification, and use \textit{balanced~accuracy~weighted} $= w_p \times \frac{\textit{TP}}{\textit{TP} + \textit{FN}} + w_n \times \frac{\textit{TN}}{\textit{TN} + \textit{FP}}$ where $w_p$ and $w_n$ respectively represent the ratio of positive relationships and the ratio negative relationships in the testing dataset (notes that $w_p + w_n = 1$), instead of the commonly used accuracy. 
\textit{balanced~accuracy~weighted} does not only show the ability of the model to predict positive relationships but also reflect its ability to predict negative relationships.

\subsection{Experimental Results and Discussions} \label{sub: results}

\subsubsection{Benchmark comparison between pretrained LM-enhanced link prediction and general machine learning models}

\begin{table*}[htb!]\footnotesize\setlength{\tabcolsep}{1.5pt}
\captionof{table}{Results for the quintuplet of \textit{(company, supplies, product, to, company)}}
\centering
\begin{tabular}{|p{2.0cm}|p{2.9cm}|p{2.9cm}|p{2.9cm}|p{2.9cm}|p{2.9cm}|}
\hline
\textbf{Country Name} & \textbf{{LogReg}} ($Acc_{bw}$/Pre/Rec) & \textbf{{LSTM}} ($Acc_{bw}$/Pre/Rec) & \textbf{CNN} ($Acc_{bw}$/Pre/Rec) & \textbf{AutoEncoder} ($Acc_{bw}$/Pre/Rec) & \textbf{ANN} ($Acc_{bw}$/Pre/Rec) \\
\hline
AUSTRALIA & 0.7953/0.7971/0.7936 & 0.8106/0.8106/0.8103 & 0.7872/0.7876/0.7873 & 0.8653/0.8675/0.8640 & \textbf{0.9037}/\textbf{0.9063}/\textbf{0.9024} \\
AUSTRIA & 0.7615/0.7619/0.7616 & 0.8853/0.8853/0.8853 & 0.9073/0.9075/0.9073 & 0.9199/0.9225/0.9200 & \textbf{0.9393}/\textbf{0.9418}/\textbf{0.9393} \\ 
BELGIUM & 0.8379/0.8375/0.8375 & 0.8748/0.8749/0.8756 & 0.8969/0.8968/0.8972 & 0.9251/0.9264/0.9267 & \textbf{0.9391}/\textbf{0.9392}/\textbf{0.9400} \\
BRAZIL & 0.8285/0.8292/0.8285 & 0.8672/0.8679/0.8672 & \textbf{0.9410}/\textbf{0.9414}/\textbf{0.9410} & 0.8899/0.8915/0.8898 & 0.9144/0.9152/0.9144 \\
CANADA & 0.8822/0.8857/0.8821 & 0.9245/0.9266/0.9246 & 0.9449/0.9458/0.9449 & 0.9702/0.9703/0.9702 & \textbf{0.9758}/\textbf{0.9759}/\textbf{0.9758} \\
CHINA & 0.8558/0.8558/0.8558 & 0.8698/0.8703/0.8697 & 0.8849/0.8852/0.8848 & 0.8841/0.8846/0.8839 & \textbf{0.8908}/\textbf{0.8909}/\textbf{0.8907} \\
FRANCE & 0.8418/0.8427/0.8417 & 0.9083/0.9093/0.9082 & 0.9265/0.9272/0.9265 & 0.9240/0.9261/0.9239 & \textbf{0.9325}/\textbf{0.9347}/\textbf{0.9324} \\
GERMANY & 0.8437/0.8466/0.8442 & 0.9062/0.9079/0.9060 & \textbf{0.9187}/\textbf{0.9189}/\textbf{0.9187} & 0.9012/0.9014/0.9012 & 0.9145/0.9146/0.9145 \\
HUNGARY & 0.8321/0.8323/0.8329 & 0.8650/0.8649/0.8656 & 0.8989/0.8988/0.8995 & 0.8820/0.8828/0.8829 & \textbf{0.8992}/\textbf{0.9000}/\textbf{0.9000} \\
INDIA & 0.8236/0.8256/0.8237 & 0.8749/0.8752/0.8749 & 0.8939/0.8940/0.8939 & 0.8736/0.8745/0.8736 & \textbf{0.9037}/\textbf{0.9039}/\textbf{0.9037} \\
INDONESIA & 0.8514/0.8522/0.8512 & 0.8686/0.8705/0.8684 & 0.9209/0.9217/0.9207 & 0.9130/0.9141/0.9128 & \textbf{0.9230}/\textbf{0.9240}/\textbf{0.9228} \\
ITALY & 0.8548/0.8554/0.8547 & 0.9138/0.9153/0.9137 & 0.9212/0.9233/0.9210 & 0.9242/0.9259/0.9241 & \textbf{0.9331}/\textbf{0.9347}/\textbf{0.9330} \\
JAPAN & 0.8469/0.8533/0.8469 & 0.8643/0.8662/0.8643 & \textbf{0.9112}/\textbf{0.9115}/\textbf{0.9112} & 0.8986/0.8990/0.8987 & 0.9080/0.9083/0.9080 \\
KOREA & 0.8782/0.8784/0.8782 & 0.9243/0.9249/0.9243 & \textbf{0.9434}/\textbf{0.9435}/\textbf{0.9434} & 0.9195/0.9195/0.9195 & 0.9311/0.9311/0.9311 \\
MALAYSIA & 0.8187/0.8206/0.8197 & 0.8645/0.8665/0.8654 & 0.9057/0.9061/0.9062 & 0.8950/0.8959/0.8956 & \textbf{0.9211}/\textbf{0.9225}/\textbf{0.9219} \\
MEXICO & 0.8513/0.8522/0.8514 & 0.9198/0.9204/0.9198 & 0.9267/0.9274/0.9266 & 0.9129/0.9132/0.9128 & \textbf{0.9361}/\textbf{0.9363}/\textbf{0.9361} \\
POLAND & 0.7763/0.7765/0.7741 & 0.8461/0.8458/0.8457 & 0.8510/0.8514/0.8497 & 0.8615/0.8619/0.8608 & \textbf{0.8760}/\textbf{0.8834}/\textbf{0.8737} \\
RUSSIA & 0.8276/0.8277/0.8272 & 0.8512/0.8523/0.8505 & 0.8726/0.8730/0.8722 & 0.8823/0.8834/0.8817 & \textbf{0.8937}/\textbf{0.8978}/\textbf{0.8935} \\
SPAIN & 0.8310/0.8351/0.8311 & 0.9030/0.9050/0.9031 & 0.9324/0.9341/0.9324 & 0.9365/0.9383/0.9365 & \textbf{0.9528}/\textbf{0.9542}/\textbf{0.9528} \\
SWEDEN & 0.8456/0.8459/0.8454 & 0.9160/0.9169/0.9157 & 0.9127/0.9157/0.9122 & 0.9543/0.9549/0.9542 & \textbf{0.9667}/\textbf{0.9674}/\textbf{0.9665} \\
TAIWAN & 0.8408/0.8408/0.8409 & 0.9219/0.9223/0.9217 & \textbf{0.9424}/\textbf{0.9424}/\textbf{0.9424} & 0.9091/0.9094/0.9089 & 0.9310/0.9311/0.9310 \\
THAILAND & 0.8614/0.8618/0.8617 & 0.8731/0.8745/0.8727 & \textbf{0.9141}/\textbf{0.9145}/\textbf{0.9139} & 0.8927/0.8928/0.8926 & 0.9055/0.9060/0.9053 \\
TURKEY & 0.8446/0.8449/0.8444 & 0.8725/0.8752/0.8718 & 0.9062/0.9075/0.9057 & 0.8975/0.8992/0.8970 & \textbf{0.9183}/\textbf{0.9192}/\textbf{0.9180} \\
U.S.A. & 0.8585/0.8603/0.8582 & 0.8859/0.8891/0.8862 & 0.9306/0.9306/0.9306 & 0.9139/0.9145/0.9138 & \textbf{0.9351}/\textbf{0.9352}/\textbf{0.9351} \\
UK & 0.8558/0.8559/0.8560 & 0.8673/0.8717/0.8680 & 0.9147/0.9148/0.9148 & 0.9073/0.9081/0.9075 & \textbf{0.9188}/\textbf{0.9191}/\textbf{0.9190} \\
\hline
\end{tabular}
\label{tab: five machine learning models without llm for supplies to}
\end{table*}

\begin{table*}[htb!]\footnotesize\setlength{\tabcolsep}{1.5pt}
\captionof{table}{Results for the quintuplet of \textit{(company, with, certificate, has, product)}}
\centering
\begin{tabular}{|p{2.0cm}|p{2.9cm}|p{2.9cm}|p{2.9cm}|p{2.9cm}|p{2.9cm}|}
\hline
\textbf{Country Name} & \textbf{{LogReg}} ($Acc_{bw}$/Pre/Rec) & \textbf{{LSTM}} ($Acc_{bw}$/Pre/Rec) & \textbf{CNN} ($Acc_{bw}$/Pre/Rec) & \textbf{AutoEncoder} ($Acc_{bw}$/Pre/Rec) & \textbf{ANN} ($Acc_{bw}$/Pre/Rec) \\
\hline
AUSTRALIA & 0.7052/0.7063/0.7052 & 0.6742/0.6750/0.6742 & 0.7052/0.7057/0.7052 & 0.7768/0.7780/0.7768 & \textbf{0.7987}/\textbf{0.8029}/\textbf{0.7987} \\
AUSTRIA & 0.6891/0.6901/0.6894 & 0.6698/0.6704/0.6695 & 0.6583/0.6622/0.6576 & 0.7708/0.7721/0.7710 & \textbf{0.8068}/\textbf{0.8072}/\textbf{0.8067} \\
BELGIUM & 0.7148/0.7162/0.7138 & 0.6784/0.6787/0.6778 & 0.7107/0.7115/0.7099 & 0.7651/0.7677/0.7642 & \textbf{0.7956}/\textbf{0.7981}/\textbf{0.7946} \\
BRAZIL & 0.7878/0.8012/0.7829 & 0.7908/0.8097/0.7852 & 0.8348/0.8361/0.8335 & 0.8185/0.8293/0.8147 & \textbf{0.8406}/\textbf{0.8440}/\textbf{0.8384} \\
CANADA & 0.7766/0.7801/0.7699 & 0.7684/0.7672/0.7683 & 0.7924/0.7911/0.7913 & 0.8225/0.8269/0.8176 & \textbf{0.8572}/\textbf{0.8585}/\textbf{0.8541} \\
CHINA & 0.7752/0.7844/0.7812 & 0.7961/0.7958/0.7971 & \textbf{0.8103}/\textbf{0.8097}/\textbf{0.8100} & 0.8058/0.8059/0.8038 & 0.8044/0.8050/0.8043 \\
FRANCE & 0.7743/0.7795/0.7716 & 0.8011/0.8032/0.7996 & 0.8183/0.8205/0.8168 & 0.8121/0.8177/0.8098 & \textbf{0.8262}/\textbf{0.8279}/\textbf{0.8249} \\
GERMANY & 0.7247/0.7248/0.7246 & 0.7610/0.7612/0.7609 & 0.7702/0.7708/0.7703 & 0.7787/0.7816/0.7781 & \textbf{0.7897}/\textbf{0.7904}/\textbf{0.7893} \\
HUNGARY & 0.6435/0.6450/0.6454 & 0.6658/0.6648/0.6649 & \textbf{0.7969}/\textbf{0.7967}/\textbf{0.7956} & 0.7353/0.7351/0.7344 & 0.7661/0.7654/0.7657 \\
INDIA & 0.7354/0.7413/0.7380 & 0.7716/0.7715/0.7712 & \textbf{0.8135}/\textbf{0.8143}/\textbf{0.8142} & 0.8067/0.8080/0.8055 & 0.8130/0.8135/0.8121 \\
INDONESIA & 0.7364/0.7409/0.7368 & 0.7278/0.7308/0.7281 & 0.7695/0.7710/0.7697 & 0.7705/0.7911/0.7713 & \textbf{0.8177}/\textbf{0.8220}/\textbf{0.8180} \\
ITALY & 0.7499/0.7546/0.7516 & 0.7283/0.7285/0.7286 & 0.8036/0.8050/0.8045 & 0.8257/0.8332/0.8278 & \textbf{0.8419}/\textbf{0.8451}/\textbf{0.8433} \\
JAPAN & 0.8413/0.8412/0.8399 & 0.8506/0.8529/0.8478 & 0.8492/0.8499/0.8474 & 0.8563/0.8584/0.8537 & \textbf{0.8590}/\textbf{0.8615}/\textbf{0.8562} \\
KOREA & 0.7466/0.7469/0.7466 & 0.7799/0.7804/0.7800 & \textbf{0.8173}/\textbf{0.8186}/\textbf{0.8173} & 0.7990/0.8046/0.7991 & 0.8039/0.8045/0.8039 \\
MALAYSIA & 0.6856/0.6864/0.6861 & 0.6937/0.6980/0.6953 & 0.7219/0.7253/0.7232 & 0.7358/0.7449/0.7377 & \textbf{0.7552}/\textbf{0.7619}/\textbf{0.7570} \\
MEXICO & 0.7342/0.7368/0.7340 & 0.6884/0.6896/0.6882 & 0.7729/0.7736/0.7729 & 0.7917/0.7982/0.7914 & \textbf{0.8214}/\textbf{0.8222}/\textbf{0.8213} \\
POLAND & 0.7339/0.7328/0.7339 & 0.7301/0.7285/0.7269 & 0.7842/0.7828/0.7828 & 0.7729/0.7725/0.7708 & \textbf{0.7954}/\textbf{0.7947}/\textbf{0.7926} \\
RUSSIA & 0.8536/0.8577/0.8488 & 0.7651/0.7640/0.7635 & 0.6365/0.6359/0.6353 & 0.8568/0.8569/0.8553 & \textbf{0.8693}/\textbf{0.8711}/\textbf{0.8663} \\
SPAIN & 0.7017/0.7037/0.7033 & 0.7592/0.7603/0.7574 & 0.7868/0.7875/0.7862 & 0.8234/0.8244/0.8232 & \textbf{0.8394}/\textbf{0.8406}/\textbf{0.8383} \\
SWEDEN & 0.6786/0.6784/0.6787 & 0.6828/0.6825/0.6817 & 0.6984/0.7007/0.6955 & 0.7531/0.7587/0.7500 & \textbf{0.8000}/\textbf{0.8057}/\textbf{0.7970} \\
TAIWAN & 0.7218/0.7221/0.7212 & 0.7639/0.7674/0.7647 & \textbf{0.8301}/\textbf{0.8309}/\textbf{0.8305} & 0.7963/0.7983/0.7970 & 0.8037/0.8043/0.8040 \\
THAILAND & 0.7064/0.7151/0.7097 & 0.7056/0.7056/0.7052 & 0.7483/0.7502/0.7496 & 0.7659/0.7666/0.7658 & \textbf{0.7824}/\textbf{0.7833}/\textbf{0.7810} \\
TURKEY & 0.7273/0.7279/0.7279 & 0.7047/0.7048/0.7039 & \textbf{0.8175}/\textbf{0.8212}/\textbf{0.8161} & 0.7796/0.7850/0.7779 & 0.8058/0.8066/0.8051 \\
U.S.A. & 0.7795/0.7863/0.7821 & 0.8148/0.8146/0.8147 & 0.8330/0.8333/0.8323 & 0.8378/0.8384/0.8372 & \textbf{0.8400}/\textbf{0.8409}/\textbf{0.8392} \\
UK & 0.7590/0.7612/0.7602 & 0.7932/0.7938/0.7934 & \textbf{0.8342}/\textbf{0.8356}/\textbf{0.8335} & 0.8117/0.8138/0.8106 & 0.8269/0.8277/0.8262 \\
\hline
\end{tabular}
\label{tab: five machine learning models without llm for make}
\end{table*}

\begin{table*}[htb!]\footnotesize\setlength{\tabcolsep}{1.5pt}
\captionof{table}{Pretrained LM-enhanced machine learning, ``all-MiniLM-L12-v2'', for different machine learning models for quintuplet \textit{(company, supplies, product, to, company)}}
\centering
\begin{tabular}{|p{2.0cm}|p{2.9cm}|p{2.9cm}|p{2.9cm}|p{2.9cm}|p{2.9cm}|}
\hline
\textbf{Country Name} & \textbf{{LogReg}} ($Acc_{bw}$/Pre/Rec) & \textbf{{LSTM}} ($Acc_{bw}$/Pre/Rec) & \textbf{CNN} ($Acc_{bw}$/Pre/Rec) & \textbf{AutoEncoder} ($Acc_{bw}$/Pre/Rec) & \textbf{ANN} ($Acc_{bw}$/Pre/Rec) \\
\hline
AUSTRALIA & \textbf{1.0000}/\textbf{1.0000}/\textbf{1.0000} & \textbf{1.0000}/\textbf{1.0000}/\textbf{1.0000} & \textbf{1.0000}/\textbf{1.0000}/\textbf{1.0000} & \textbf{1.0000}/\textbf{1.0000}/\textbf{1.0000} & 0.9991/0.9990/0.9991 \\
AUSTRIA & \textbf{1.0000}/\textbf{1.0000}/\textbf{1.0000} & \textbf{1.0000}/\textbf{1.0000}/\textbf{1.0000} & \textbf{1.0000}/\textbf{1.0000}/\textbf{1.0000} & \textbf{1.0000}/\textbf{1.0000}/\textbf{1.0000} & 0.9614/0.9658/0.9615 \\
BELGIUM & \textbf{1.0000}/\textbf{1.0000}/\textbf{1.0000} & \textbf{1.0000}/\textbf{1.0000}/\textbf{1.0000} & \textbf{1.0000}/\textbf{1.0000}/\textbf{1.0000} & \textbf{1.0000}/\textbf{1.0000}/\textbf{1.0000} & 0.9982/0.9981/0.9982 \\
BRAZIL & \textbf{1.0000}/\textbf{1.0000}/\textbf{1.0000} & 0.9990/0.9990/0.9990 & 0.9996/0.9996/0.9996 & 0.9997/0.9997/0.9997 & 0.9994/0.9994/0.9994 \\
CANADA & \textbf{1.0000}/\textbf{1.0000}/\textbf{1.0000} & \textbf{1.0000}/\textbf{1.0000}/\textbf{1.0000} & \textbf{1.0000}/\textbf{1.0000}/\textbf{1.0000} & \textbf{1.0000}/\textbf{1.0000}/\textbf{1.0000} & 0.9998/0.9998/0.9998 \\
CHINA & \textbf{1.0000}/\textbf{1.0000}/\textbf{1.0000} & 0.9999/0.9999/0.9999 & 0.9998/0.9998/0.9998 & 0.9486/0.9239/0.9492 & 0.9999/0.9999/0.9999 \\
FRANCE & \textbf{1.0000}/\textbf{1.0000}/\textbf{1.0000} & \textbf{1.0000}/\textbf{1.0000}/\textbf{1.0000} & 0.9998/0.9998/0.9998 & 0.9999/0.9999/0.9999 & 0.9989/0.9989/0.9989 \\
GERMANY & \textbf{1.0000}/\textbf{1.0000}/\textbf{1.0000} & \textbf{1.0000}/\textbf{1.0000}/\textbf{1.0000} & \textbf{1.0000}/\textbf{1.0000}/\textbf{1.0000} & 0.9999/0.9999/0.9999 & 0.9998/0.9998/0.9998 \\
HUNGARY & \textbf{1.0000}/\textbf{1.0000}/\textbf{1.0000} & 0.9985/0.9985/0.9986 & 0.9998/0.9998/0.9998 & \textbf{1.0000}/\textbf{1.0000}/\textbf{1.0000} & 0.9989/0.9988/0.9989 \\
INDIA & \textbf{1.0000}/\textbf{1.0000}/\textbf{1.0000} & \textbf{1.0000}/\textbf{1.0000}/\textbf{1.0000} & \textbf{1.0000}/\textbf{1.0000}/\textbf{1.0000} & 0.9993/0.9993/0.9993 & 0.9999/0.9999/0.9999 \\
INDONESIA & \textbf{1.0000}/\textbf{1.0000}/\textbf{1.0000} & \textbf{1.0000}/\textbf{1.0000}/\textbf{1.0000} & \textbf{1.0000}/\textbf{1.0000}/\textbf{1.0000} & \textbf{1.0000}/\textbf{1.0000}/\textbf{1.0000} & 0.9983/0.9983/0.9983 \\
ITALY & \textbf{1.0000}/\textbf{1.0000}/\textbf{1.0000} & \textbf{1.0000}/\textbf{1.0000}/\textbf{1.0000} & 0.9998/0.9998/0.9998 & \textbf{1.0000}/\textbf{1.0000}/\textbf{1.0000} & 0.9999/0.9999/0.9999 \\
JAPAN & \textbf{1.0000}/\textbf{1.0000}/\textbf{1.0000} & 0.9999/0.9999/0.9999 & 0.9999/0.9999/0.9999 & 0.9980/0.9980/0.9980 & 0.9998/0.9998/0.9998 \\
KOREA & \textbf{1.0000}/\textbf{1.0000}/\textbf{1.0000} & \textbf{1.0000}/\textbf{1.0000}/\textbf{1.0000} & \textbf{1.0000}/\textbf{1.0000}/\textbf{1.0000} & \textbf{1.0000}/\textbf{1.0000}/\textbf{1.0000} & 0.9999/0.9999/0.9999 \\
MALAYSIA & \textbf{1.0000}/\textbf{1.0000}/\textbf{1.0000} & \textbf{1.0000}/\textbf{1.0000}/\textbf{1.0000} & 0.9997/0.9997/0.9997 & 0.9999/0.9999/0.9999 & 0.9959/0.9960/0.9958 \\
MEXICO & \textbf{1.0000}/\textbf{1.0000}/\textbf{1.0000} & \textbf{1.0000}/\textbf{1.0000}/\textbf{1.0000} & \textbf{1.0000}/\textbf{1.0000}/\textbf{1.0000} & \textbf{1.0000}/\textbf{1.0000}/\textbf{1.0000} & 0.9999/0.9999/0.9999 \\
POLAND & \textbf{1.0000}/\textbf{1.0000}/\textbf{1.0000} & \textbf{1.0000}/\textbf{1.0000}/\textbf{1.0000} & \textbf{1.0000}/\textbf{1.0000}/\textbf{1.0000} & \textbf{1.0000}/\textbf{1.0000}/\textbf{1.0000} & 0.9979/0.9979/0.9979 \\
RUSSIA & \textbf{1.0000}/\textbf{1.0000}/\textbf{1.0000} & \textbf{1.0000}/\textbf{1.0000}/\textbf{1.0000} & 0.9997/0.9997/0.9996 & \textbf{1.0000}/\textbf{1.0000}/\textbf{1.0000} & 0.9852/0.9881/0.9856 \\
SPAIN & \textbf{1.0000}/\textbf{1.0000}/\textbf{1.0000} & \textbf{1.0000}/\textbf{1.0000}/\textbf{1.0000} & \textbf{1.0000}/\textbf{1.0000}/\textbf{1.0000} & \textbf{1.0000}/\textbf{1.0000}/\textbf{1.0000} & 0.9994/0.9994/0.9994 \\
SWEDEN & \textbf{1.0000}/\textbf{1.0000}/\textbf{1.0000} & \textbf{1.0000}/\textbf{1.0000}/\textbf{1.0000} & \textbf{1.0000}/\textbf{1.0000}/\textbf{1.0000} & \textbf{1.0000}/\textbf{1.0000}/\textbf{1.0000} & 0.9999/0.9999/0.9999 \\
TAIWAN & \textbf{1.0000}/\textbf{1.0000}/\textbf{1.0000} & 0.9995/0.9995/0.9995 & \textbf{1.0000}/\textbf{1.0000}/\textbf{1.0000} & \textbf{1.0000}/\textbf{1.0000}/\textbf{1.0000} & 0.9992/0.9992/0.9992 \\
THAILAND & \textbf{1.0000}/\textbf{1.0000}/\textbf{1.0000} & \textbf{1.0000}/\textbf{1.0000}/\textbf{1.0000} & \textbf{1.0000}/\textbf{1.0000}/\textbf{1.0000} & \textbf{1.0000}/\textbf{1.0000}/\textbf{1.0000} & 0.9999/0.9999/0.9999 \\
TURKEY & \textbf{1.0000}/\textbf{1.0000}/\textbf{1.0000} & 0.9999/0.9999/0.9999 & \textbf{1.0000}/\textbf{1.0000}/\textbf{1.0000} & \textbf{1.0000}/\textbf{1.0000}/\textbf{1.0000} & 0.9999/0.9999/0.9999 \\
U.S.A. & \textbf{1.0000}/\textbf{1.0000}/\textbf{1.0000} & 0.9998/0.9998/0.9998 & 0.9997/0.9997/0.9997 & 0.9998/0.9998/0.9998 & 0.9998/0.9998/0.9998 \\
UK & \textbf{1.0000}/\textbf{1.0000}/\textbf{1.0000} & 0.9995/0.9995/0.9995 & \textbf{1.0000}/\textbf{1.0000}/\textbf{1.0000} & 0.9999/0.9999/0.9999 & 0.9994/0.9994/0.9994 \\
\hline
\end{tabular}
\label{tab: five machine learning models with llm for supplies to}
\end{table*}

\begin{table*}[htb!]\footnotesize\setlength{\tabcolsep}{1.5pt}
\captionof{table}{Pretrained LM-enhanced machine learning, ``all-MiniLM-L12-v2'', for different machine learning models to predict relationships in the quintuplet of \textit{(company, with, certificate, has, product)}}
\centering
\begin{tabular}{|p{2.0cm}|p{2.9cm}|p{2.9cm}|p{2.9cm}|p{2.9cm}|p{2.9cm}|}
\hline
\textbf{Country Name} & \textbf{{LogReg}} ($Acc_{bw}$/Pre/Rec) & \textbf{{LSTM}} ($Acc_{bw}$/Pre/Rec) & \textbf{CNN} ($Acc_{bw}$/Pre/Rec) & \textbf{AutoEncoder} ($Acc_{bw}$/Pre/Rec) & \textbf{ANN} ($Acc_{bw}$/Pre/Rec) \\
\hline
AUSTRALIA & \textbf{1.0000}/\textbf{1.0000}/\textbf{1.0000} & \textbf{1.0000}/\textbf{1.0000}/\textbf{1.0000} & \textbf{1.0000}/\textbf{1.0000}/\textbf{1.0000} & \textbf{1.0000}/\textbf{1.0000}/\textbf{1.0000} & \textbf{1.0000}/\textbf{1.0000}/\textbf{1.0000} \\
AUSTRIA & \textbf{1.0000}/\textbf{1.0000}/\textbf{1.0000} & 0.9990/0.9990/0.9989 & 0.9990/0.9990/0.9989 & \textbf{1.0000}/\textbf{1.0000}/\textbf{1.0000} & \textbf{1.0000}/\textbf{1.0000}/\textbf{1.0000} \\
BELGIUM & \textbf{1.0000}/\textbf{1.0000}/\textbf{1.0000} & \textbf{1.0000}/\textbf{1.0000}/\textbf{1.0000} & \textbf{1.0000}/\textbf{1.0000}/\textbf{1.0000} & 0.9992/0.9992/0.9992 & 0.9984/0.9984/0.9985 \\
BRAZIL & \textbf{1.0000}/\textbf{1.0000}/\textbf{1.0000} & 0.9899/0.9899/0.9900 & 0.9984/0.9984/0.9983 & 0.9992/0.9992/0.9992 & 0.9995/0.9995/0.9995 \\
CANADA & \textbf{1.0000}/\textbf{1.0000}/\textbf{1.0000} & \textbf{1.0000}/\textbf{1.0000}/\textbf{1.0000} & \textbf{1.0000}/\textbf{1.0000}/\textbf{1.0000} & 0.9998/0.9998/0.9998 & 0.9990/0.9990/0.9990 \\
CHINA & \textbf{1.0000}/\textbf{1.0000}/\textbf{1.0000} & 0.9996/0.9996/0.9996 & 0.9992/0.9992/0.9992 & 0.9992/0.9992/0.9992 & 0.9998/0.9998/0.9998 \\
FRANCE & \textbf{1.0000}/\textbf{1.0000}/\textbf{1.0000} & 0.9995/0.9995/0.9995 & \textbf{1.0000}/\textbf{1.0000}/\textbf{1.0000} & 0.9998/0.9998/0.9998 & 0.9996/0.9996/0.9996 \\
GERMANY & \textbf{1.0000}/\textbf{1.0000}/\textbf{1.0000} & 0.9994/0.9994/0.9994 & 0.9996/0.9996/0.9996 & 0.9998/0.9998/0.9998 & 0.9994/0.9994/0.9994 \\
HUNGARY & \textbf{1.0000}/\textbf{1.0000}/\textbf{1.0000} & \textbf{1.0000}/\textbf{1.0000}/\textbf{1.0000} & \textbf{1.0000}/\textbf{1.0000}/\textbf{1.0000} & \textbf{1.0000}/\textbf{1.0000}/\textbf{1.0000} & \textbf{1.0000}/\textbf{1.0000}/\textbf{1.0000} \\
INDIA & \textbf{1.0000}/\textbf{1.0000}/\textbf{1.0000} & \textbf{1.0000}/\textbf{1.0000}/\textbf{1.0000} & \textbf{1.0000}/\textbf{1.0000}/\textbf{1.0000} & 0.9997/0.9997/0.9997 & 0.9997/0.9996/0.9997 \\
INDONESIA & \textbf{1.0000}/\textbf{1.0000}/\textbf{1.0000} & \textbf{1.0000}/\textbf{1.0000}/\textbf{1.0000} & \textbf{1.0000}/\textbf{1.0000}/\textbf{1.0000} & \textbf{1.0000}/\textbf{1.0000}/\textbf{1.0000} & \textbf{1.0000}/\textbf{1.0000}/\textbf{1.0000} \\
ITALY & \textbf{1.0000}/\textbf{1.0000}/\textbf{1.0000} & 0.9993/0.9993/0.9993 & 0.9999/0.9999/0.9999 & \textbf{1.0000}/\textbf{1.0000}/\textbf{1.0000} & 0.9987/0.9987/0.9987 \\
JAPAN & 0.9998/0.9998/0.9999 & 0.9998/0.9998/0.9999 & 0.9999/0.9999/0.9999 & 0.9993/0.9993/0.9994 & 0.9992/0.9992/0.9993 \\
KOREA & \textbf{1.0000}/\textbf{1.0000}/\textbf{1.0000} & \textbf{1.0000}/\textbf{1.0000}/\textbf{1.0000} & \textbf{1.0000}/\textbf{1.0000}/\textbf{1.0000} & \textbf{1.0000}/\textbf{1.0000}/\textbf{1.0000} & \textbf{1.0000}/\textbf{1.0000}/\textbf{1.0000} \\
MALAYSIA & 0.9984/0.9985/0.9984 & 0.9984/0.9984/0.9985 & \textbf{1.0000}/\textbf{1.0000}/\textbf{1.0000} & 0.9984/0.9985/0.9984 & 0.9970/0.9971/0.9970 \\
MEXICO & 0.9983/0.9983/0.9983 & \textbf{1.0000}/\textbf{1.0000}/\textbf{1.0000} & 0.9998/0.9998/0.9998 & 0.9988/0.9988/0.9988 & 0.9979/0.9979/0.9979 \\
POLAND & \textbf{1.0000}/\textbf{1.0000}/\textbf{1.0000} & \textbf{1.0000}/\textbf{1.0000}/\textbf{1.0000} & 0.9983/0.9982/0.9985 & 0.9993/0.9993/0.9994 & 0.9992/0.9992/0.9992 \\
RUSSIA & \textbf{1.0000}/\textbf{1.0000}/\textbf{1.0000} & \textbf{1.0000}/\textbf{1.0000}/\textbf{1.0000} & 0.9990/0.9990/0.9989 & \textbf{1.0000}/\textbf{1.0000}/\textbf{1.0000} & \textbf{1.0000}/\textbf{1.0000}/\textbf{1.0000} \\
SPAIN & \textbf{1.0000}/\textbf{1.0000}/\textbf{1.0000} & \textbf{1.0000}/\textbf{1.0000}/\textbf{1.0000} & 0.9991/0.9991/0.9992 & 0.9993/0.9993/0.9993 & 0.9995/0.9995/0.9995 \\
SWEDEN & 0.9984/0.9984/0.9985 & 0.9984/0.9984/0.9985 & 0.9995/0.9995/0.9995 & 0.9979/0.9980/0.9979 & 0.9984/0.9984/0.9985 \\
TAIWAN & \textbf{1.0000}/\textbf{1.0000}/\textbf{1.0000} & 0.9991/0.9991/0.9991 & 0.9997/0.9997/0.9997 & \textbf{1.0000}/\textbf{1.0000}/\textbf{1.0000} & 0.9995/0.9995/0.9995 \\
THAILAND & \textbf{1.0000}/\textbf{1.0000}/\textbf{1.0000} & 0.9999/0.9999/0.9999 & \textbf{1.0000}/\textbf{1.0000}/\textbf{1.0000} & \textbf{1.0000}/\textbf{1.0000}/\textbf{1.0000} & 0.9994/0.9994/0.9994 \\
TURKEY & \textbf{1.0000}/\textbf{1.0000}/\textbf{1.0000} & \textbf{1.0000}/\textbf{1.0000}/\textbf{1.0000} & \textbf{1.0000}/\textbf{1.0000}/\textbf{1.0000} & \textbf{1.0000}/\textbf{1.0000}/\textbf{1.0000} & 0.9999/0.9999/0.9999 \\
U.S.A. & \textbf{1.0000}/\textbf{1.0000}/\textbf{1.0000} & 0.9994/0.9994/0.9994 & 0.9998/0.9998/0.9998 & 0.9998/0.9998/0.9998 & 0.9995/0.9995/0.9995 \\
UK & \textbf{1.0000}/\textbf{1.0000}/\textbf{1.0000} & 0.9989/0.9989/0.9989 & 0.9993/0.9993/0.9993 & \textbf{1.0000}/\textbf{1.0000}/\textbf{1.0000} & 0.9997/0.9997/0.9998 \\
\hline
\end{tabular}
\label{tab: five machine learning models with llm for make}
\end{table*}

Table~\ref{tab: five machine learning models without llm for supplies to} and Table~\ref{tab: five machine learning models without llm for make} respectively show results achieved by the machine learning models to predict the quintuplets of \textit{(company, supplies, product, to, company)} and \textit{(company, with, certificate, has, product)}, while Table~\ref{tab: five machine learning models with llm for supplies to} and Table~\ref{tab: five machine learning models with llm for make} present results achieved by the pretrained LM-enhanced approach.

\begin{figure*}[hbt!]
    \centering
   \subfloat[]{\includegraphics[width=0.85\textwidth]{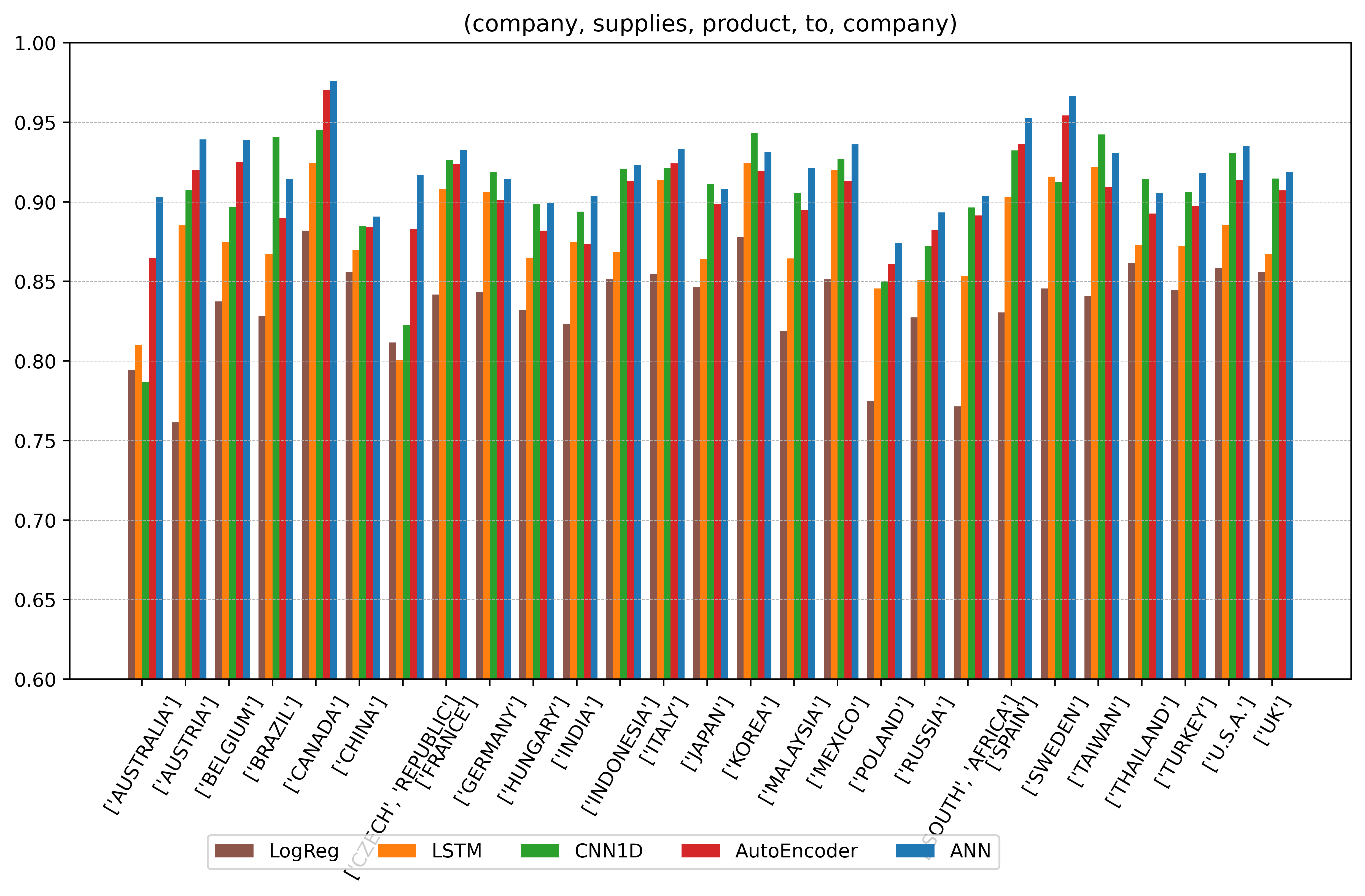}}
    \vspace{-0cm}
    \subfloat[]{\includegraphics[width=0.85\textwidth]{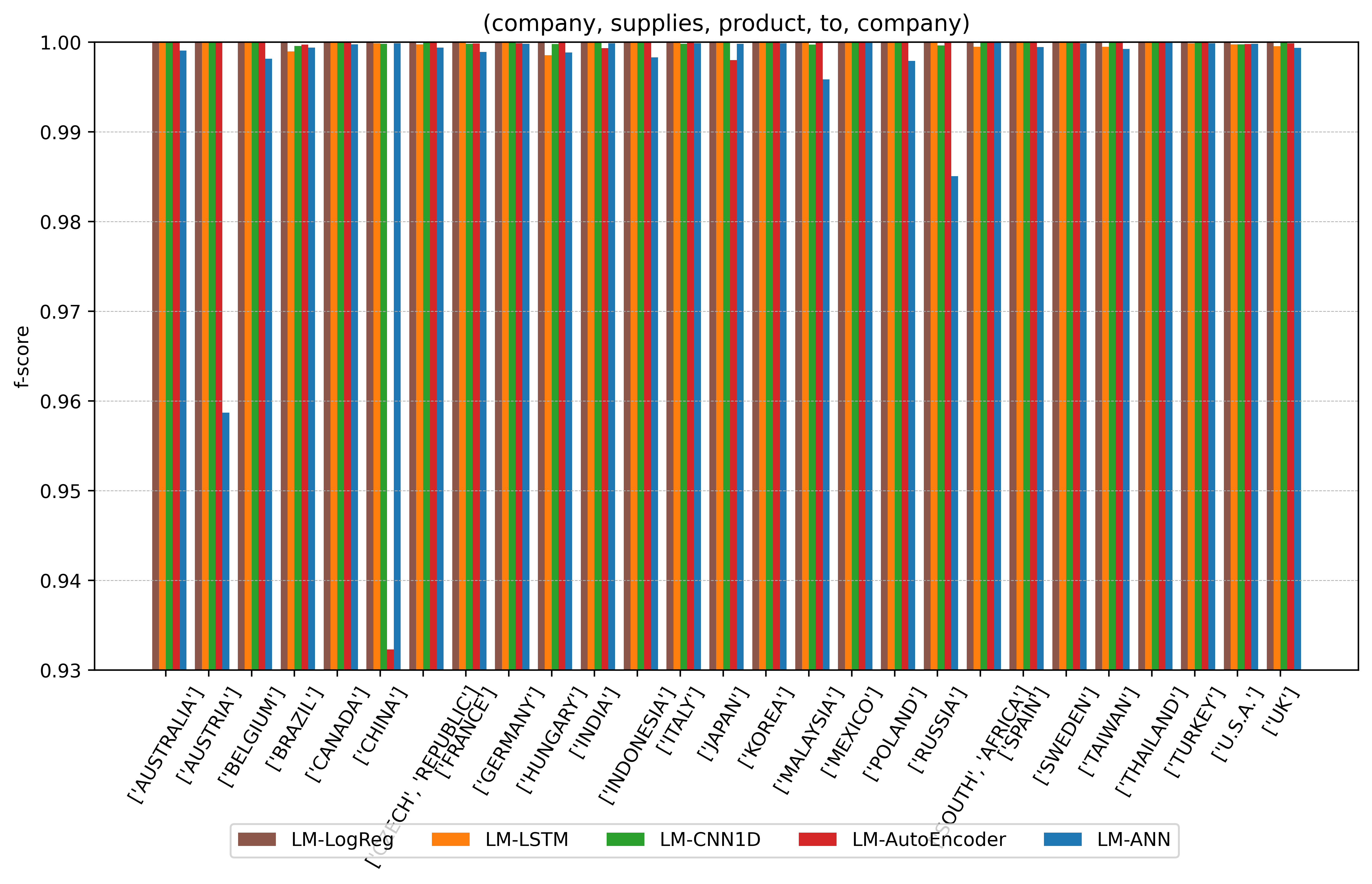}}
    \vspace{-0.2cm}
    \caption{(a) results on quintuplet  \textit{(company, supplies, product, to, company)} achieved by different machine learning models  (b) results achieved by pretrained LM-enhanced machine learning models with ``all-MiniLM-L12-v2'' \newline Figure 3 Alt-text: The diagram shows results on quintuplet \textit{(company, supplies, product, to, company)} achieved by different machine learning models and also the results achieved by pretrained LM-enhanced machine learning models with ``all-MiniLM-L12-v2''.}
    \label{fig: supply result}
\end{figure*}

\begin{figure*}[hbt!]
    \centering
   \subfloat[]{\includegraphics[width=0.85\textwidth]{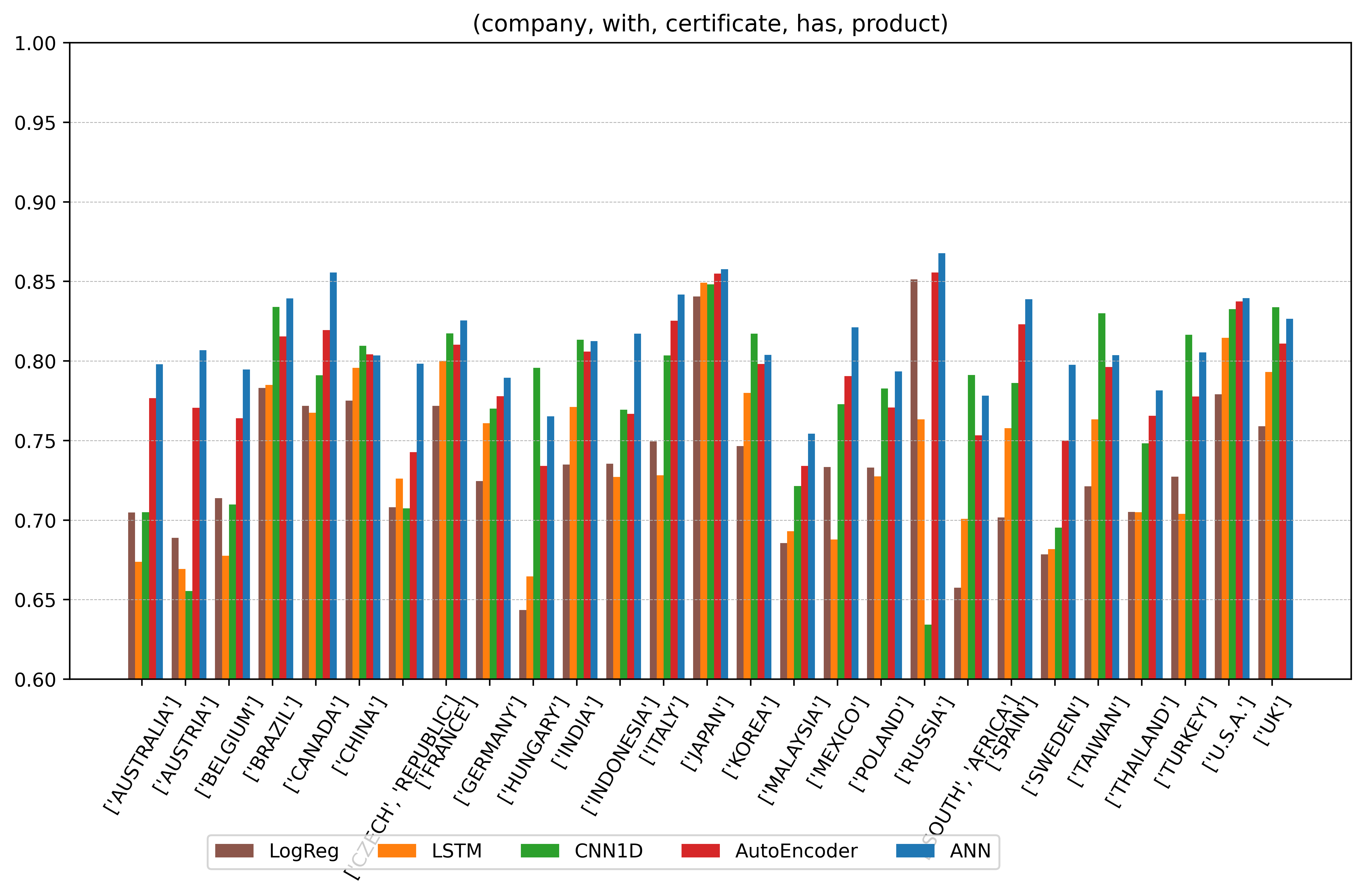}}
    \vspace{-0cm}
    \subfloat[]{\includegraphics[width=0.85\textwidth]{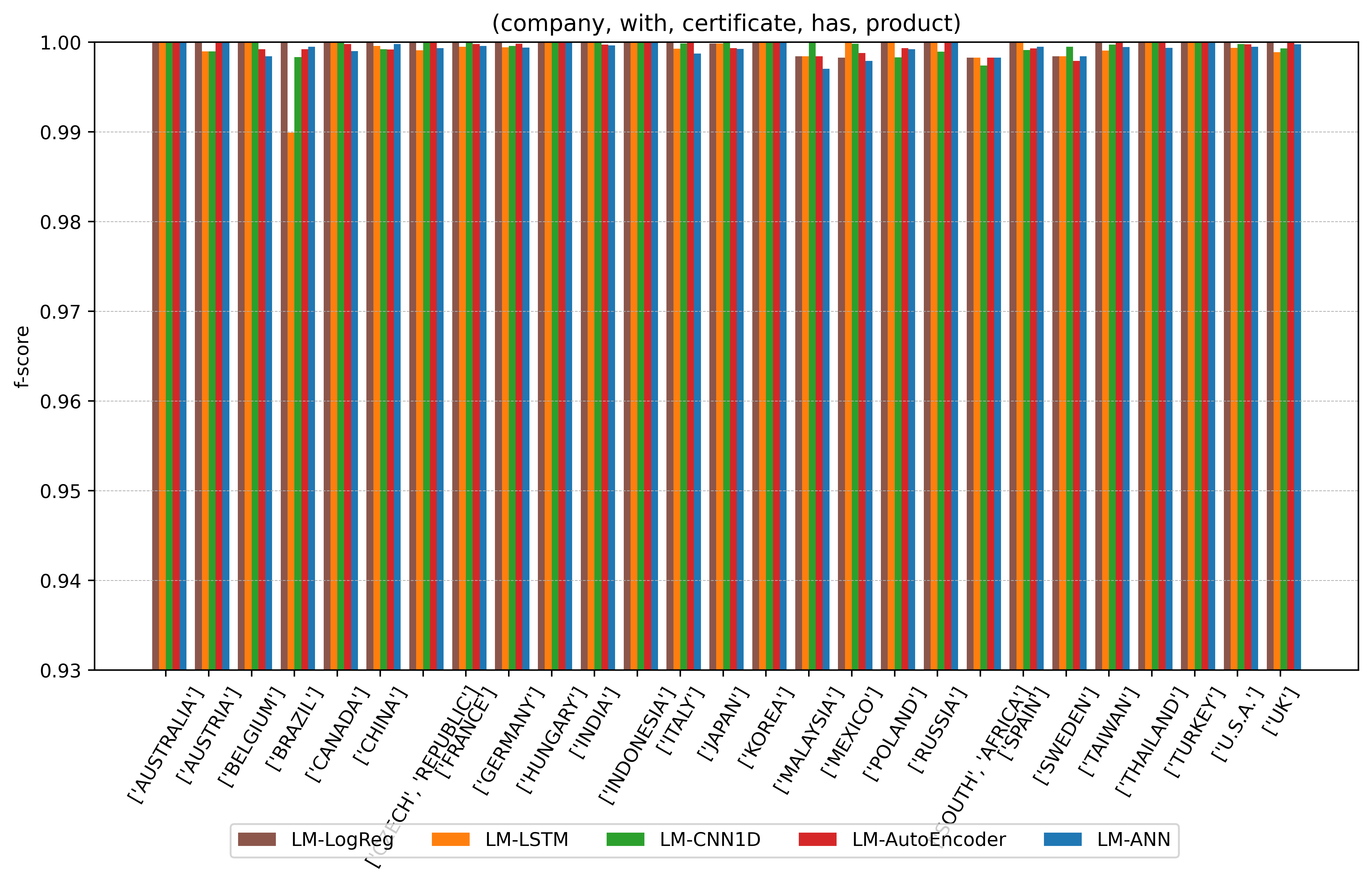}}
    \vspace{-0.2cm}
    \caption{(a) shows results of predicting relationships in the quintuplet of \textit{(company, with, certificate, has, product)} achieved by five machine learning models on all countries' datasets while (b) presents the results achieved by our proposed approach using five machine learning models empowered by the pretrained LM, ``all-MiniLM-L12-v2''. \newline
    Figure 4 Alt-text: The diagram presents results of predicting relationships in the quintuplet of \textit{(company, with, certificate, has, product)} achieved by five machine learning models on all countries' datasets, and also the results achieved by our proposed approach using five machine learning models empowered by the pretrained LM, ``all-MiniLM-L12-v2''.}
    \label{fig: make result}
\end{figure*}

Based on balanced accuracy weighted ($Acc_{bw}$), precision (Pre) and recall (Rec) in these tables, an obvious finding is that the pretrained LM-enhanced link prediction outperforms all general machine learning models for both quintuplets on all datasets. 
This finding can also be observed in Figure~\ref{fig: supply result} and Figure~\ref{fig: make result} that show the $fw-score$ achieved by our proposed approach and general machine learning models for the link predictions in \textit{(company, supplies, product, to, company)} and \textit{(company, with, certificate, has, product)}. 
These results indicate that pretrained LMs indeed can help general machine learning models achieve better link predictions in supply chain networks, confirming observation that machine learning models benefit from the relational knowledge learned in pretrained LMs \citep{bouraoui2020inducing, petroni2019language, safavi2021relational}.

In addition, the pretrained LM-enhanced approach presents more consistent results across different quintuplets, compared to results obtained by direct application of machine learning models, which yield poor performance in predicting quintuplets of \textit{(company, with, certificate, has, product)} compared to quintuplets of \textit{(company, supplies, product, to, company)}.

\subsubsection{Comparisons between Machine Learning Models}

Based on results achieved by machine learning models shown in Table~\ref{tab: five machine learning models without llm for supplies to}, Table~\ref{tab: five machine learning models without llm for make}, Figure~\ref{fig: supply result} (a) and Figure~\ref{fig: make result} (a), we find that, on all datasets, all machine learning models perform better for predicting relationships in the quintuplet of \textit{(company, supplies, product, to, company)} than in \textit{(company, with, certificate, has, product)}. 
This is because the number of \textit{(company, supplies, product, to, company)} samples is more than the samples of \textit{(company, with, certificate, has, product)} in each country's dataset, which can be indirectly observed by a large number of companies and a small number of certificates and products shown in Table~\ref{tab: marklines data}.

Among the different machine learning models, ANN, CNN1D and AutoEncoder predict relationships more accurately in both types of quintuplets than LSTM and LogReg. 
This finding matches a common conclusion from many existing works \citep{zheng2023federated, caruana2006empirical, abu2007comparison} that show ANN and CNN are better than LogReg and LSTM in binary classification tasks.

\subsubsection{Comparisons between Pretrained Language Models}

\begin{figure*}[hbt!]
    \centering
   \subfloat[]{\includegraphics[width=0.85\textwidth]{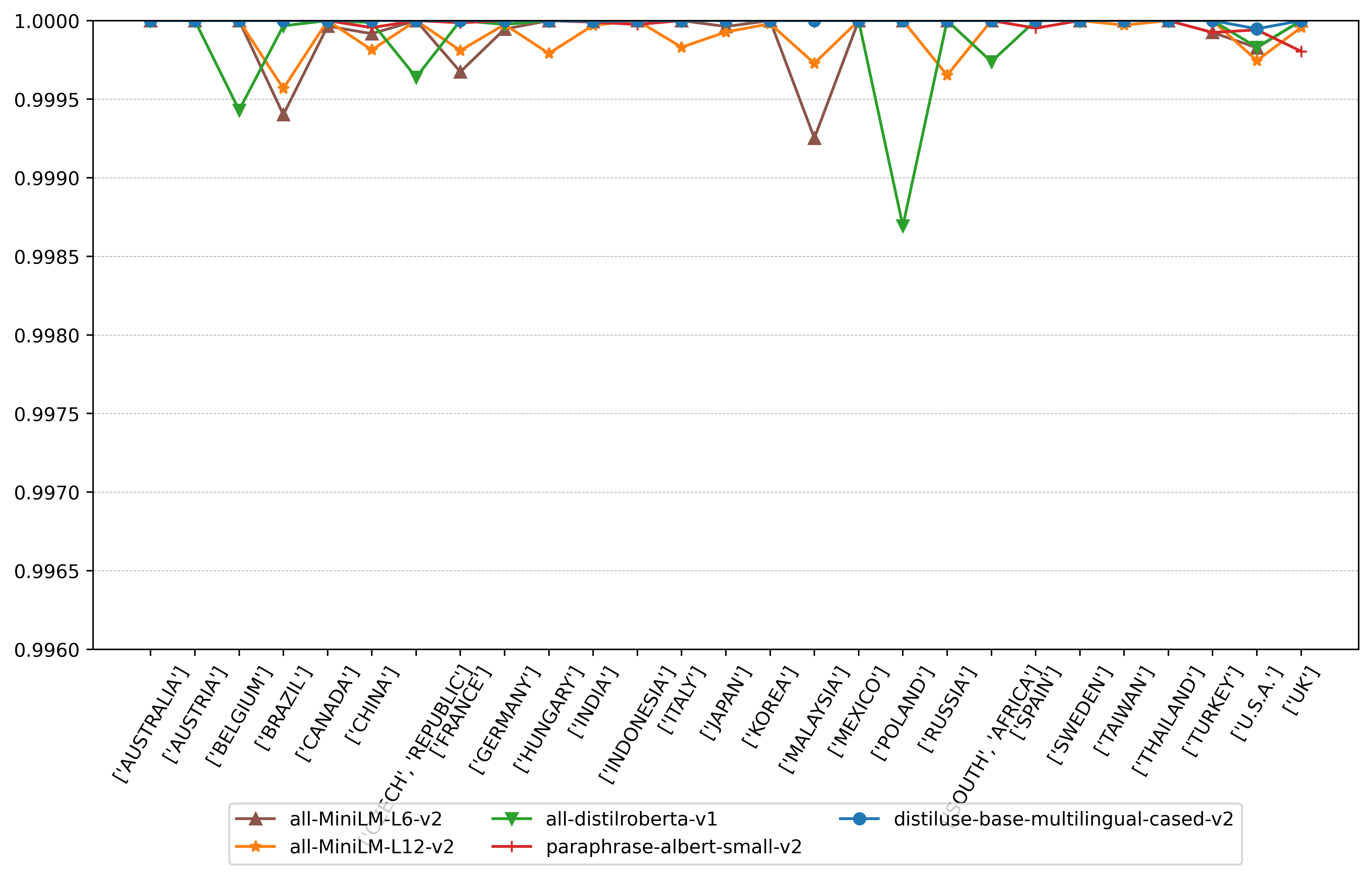}}
    \vspace{-0cm}
    \subfloat[]{\includegraphics[width=0.85\textwidth]{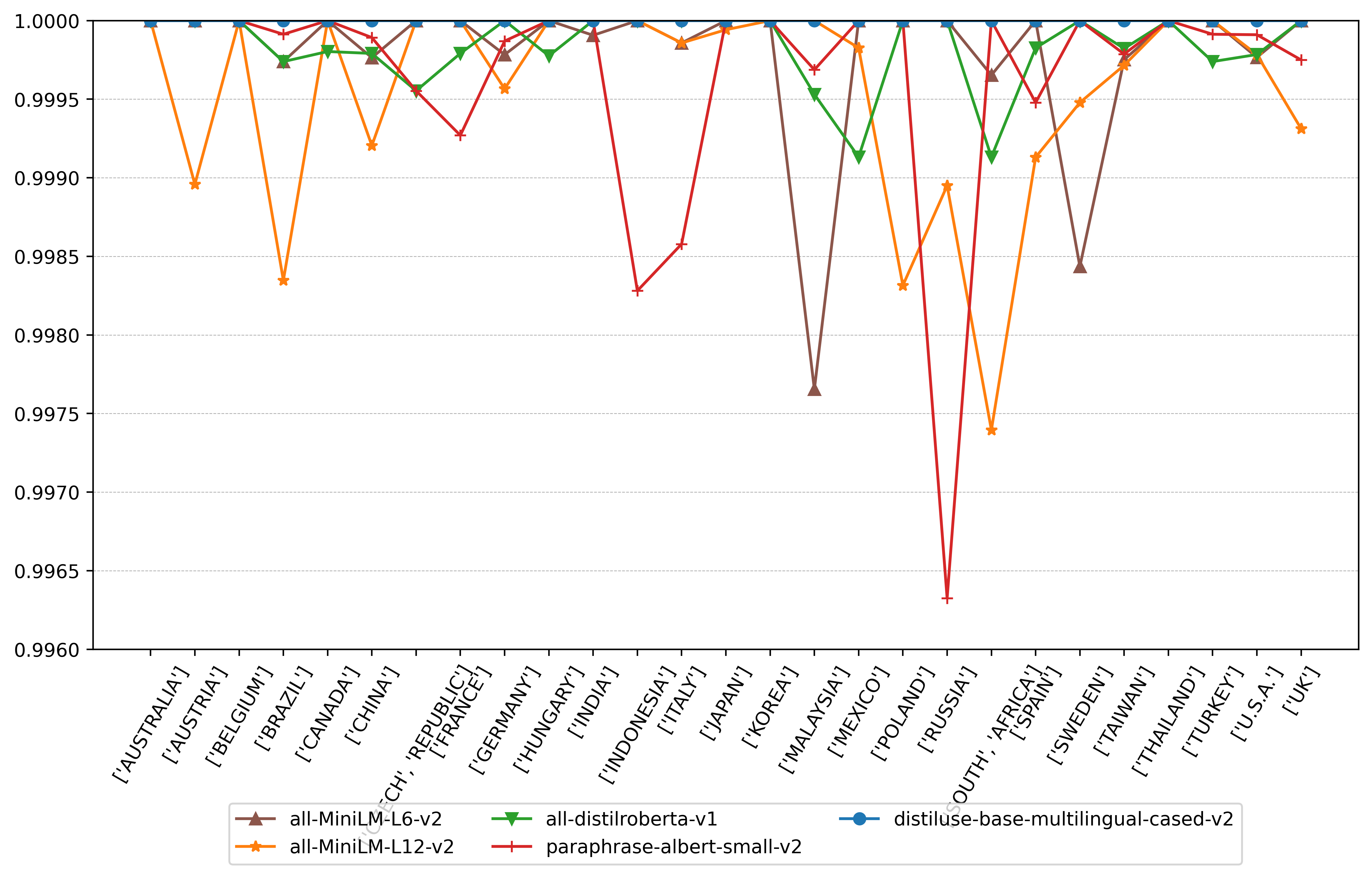}}
    \vspace{-0.2cm}
    \caption{(a) results of predicting relationships in a quintruplet of \textit{(company, supplies, product, to, company)} achieved by pretrained LM-enhanced CNN  (b) results of predicting relationships in quintuplet \textit{(company, with, certificate, has, product)}. \newline
    Figure 5 Alt-text: The diagram shows results of predicting relationships in a quintuplet of \textit{(company, supplies, product, to, company)} achieved by pretrained LM-enhanced CNN, and also the results of predicting relationships in quintuplet \textit{(company, with, certificate, has, product)}.}
    \label{fig: five nlp results}
\end{figure*}

Figure~\ref{fig: five nlp results} shows that results achieved by the pretrained LMs to enhance the CNN1D model provide higher prediction accuracy on both types of quintuplets. 
Although all five pretrained LMs improve relationship prediction accuracy, the performances of pretrained LMs have subtle differences. 

For example, notice that in Figure~\ref{fig: five nlp results} (a) , ``distiluse-base-multilingual-cased-v2'' outperforms all other four in predicting \textit{(company, supplies, product, to, company)}. 
This can also be observed in Figure~\ref{fig: five nlp results} (b) for predicting quintuplet \textit{(company, with, certificate, has, product)}.

The performance of ``distiluse-base-multilingual-cased-v2'' is more consistent across all tasks, compared to the other four models. 
This pretrained LM is trained on data in 15 languages \citep{reimers-2019-sentence-bert} providing richer knowledge hidden in different languages. 
Particularly for using language models as knowledge bases, language models trained on multilingual data can learn better representations than being trained on monolingual data \citep{pratap2020massively, kassner2021multilingual}.

Regarding the relationship predictions in different types of quintuplets, four of five pretrained LMs excluding ``distiluse-base-multilingual-cased-v2'' perform better predictions on \textit{(company, supplies, product, to, company)} than on \textit{(company, with, certificate, has, product)}. 
This is because relationships in the former describe a type of network-level information in supply chains while relationships in the latter present a type of internal information in a company. 
Network-level information in supply chains, such as relationships between companies, tends to be more accessible and observable compared to internal information of a specific company, due to network-level information that can be inferred from public sources and industry publications. 
Internal information about a company such as specific product certificates or product processes is more sensitive and not readily available to external observers.

\subsubsection{Summary of Findings}

Our findings can be summarised as follows:

\begin{itemize}
    \item Enhancing link prediction models with pretrained LMs outperforms all five benchmarks, indicating that pretrained LMs indeed can help common machine learning models achieve better relationship predictions in supply chain networks due to learned relational knowledge in these pretrained LMs. 

    \item Pretrained LM-enhanced machine learning model on prediction tasks is more consistent than using machine learning model alone, and is less affected by differences in dataset size.

    \item Pretrained LM-enhanced models are better in predicting relationships that rely on network-level information, compared to relationships that rely on internal-company information. 
    This is because network-level information in supply chains tends to be more accessible and observable from public sources, compared to internal information of a specific entity in the network. 
    As such, pretrained LM-enhancement works better in cases where we predict who supplies, which product to whom, compared to for example, which quality certification a company may have for which product.

    \item ANN, CNN1D and AutoEncoder that are commonly good at solving binary classification problems can predict relationships in supply chain networks more accurately than LSTM and LogReg in our case. 

    \item Pretrained multilingual LMs benefit common machine learning models better than monolingual LMs due to being trained on multilingual data to learn better representations.  
    
\end{itemize}


\section{Conclusions, Managerial Implications, and Future Works} \label{sec: conclusion}

Relationship prediction, also called link prediction, or supply network reconstruction, is an emergent area of ``digital supply chain surveillance'' research that aims to increase visibility of supply networks using data-driven techniques without having to rely on the willingness of supply chain actors to share information. 
Although many of the proposed methods have been very successful in reconstructing supply-buy relationships the context in which these relationships are embedded has thus far lacked attention. 
This hinders researchers and practitioners to take full advantage of these methods, as they cannot accurately differentiate between a transactional relationship and established supply relationships that characterise physical resources needed to produce a product. 
As such, estimations of resilience, distance to malicious actors and harmful practices based remain inaccurate.

Recently, Generative AI (GenAI) methods such as LLMs have become popular in eliciting information patterns from natural language data. There is also much hype in their potential in SCM. However we cannot simply ask an LLM whether a supply relationship exists, due to their hallucination problem. Hence we need methods to combine the power of GenAI methods with structured, guaranteeable methods when it comes to supply network surveillance. To date, there have been no studies on the use of LLMs for supply network surveillance. 

In this work, we developed a framework that used GenAI and machine learning for predicting complex relationships of entities in supply chain networks. 
We defined a new term, ``\textit{quintuplet}'', to describe the complex relationships that consist of multiple connected relationships and present a complete information flow, instead of commonly used triplets in the literature describing partial information. 
Quintuplets provide contextual information to relationships, such as the flow of products a relationship is set for. 
We define the link prediction problem as a binary classification problem, aiming to predict whether a quintuplet exists or not. 
We then develop a machine learning-based approach enhanced by a type of GenAI models, pretrained LLMs, used as a knowledge base. This allows us to tap into the benefits of LLMs while at the same time, mitigate hallucinations.

A practical case study is used to evaluate the proposed approach with comparative benchmarks that use machine learning methods without pretained LM enhancement. 
Pretrained LM-enhanced quintuplet prediction surpasses all benchmarks and provides consistent performance across all datasets with the advantage of providing contextual information, allowing stakeholders to be able to track the movements of products in a global network.

A secondary contribution of our research is a practical use case demonstrating the potential of GenAI models, particularly pretrained LMs, in supply chain management. 
Most works in the literature regarding the use of GenAI models like large language models in supply chains are limited to theoretical discussion. 
Therefore, our work here, as a practical study, helps bridge the gap between academic discourse and real-world application, and opens a door for more practical works in solving supply chain challenges with the use of language models.

Lastly, we have shows that pretrained LMs contain currently untapped relational supply chain knowledge presenting opportunities \citep{petroni2019language, bouraoui2020inducing}. 
This finding should encourage researchers to explore further uses of pretrained LMs for solving other types of supply chain challenges. 
In future work, we aim to test the developed link prediction method to other use case in industries, and explore other types of contextual knowledge that can be gained from language models for supply chain management.

\section*{Data Availability Statement}

Due to the commercially sensitive nature of this research, supporting data is not available.

{\small
	\bibliographystyle{chicago}
	\bibliography{reference.bib}
}



\end{document}